\documentclass[10pt,final,superscriptaddress,english,twocolumn,amssymb,aps,prb,
longbibliography]{revtex4-2}
\usepackage{graphicx}
\usepackage{natbib}
\usepackage{physics}
\usepackage{amsmath}
\usepackage{amssymb}
\usepackage{appendix}
\usepackage{booktabs}

\usepackage{bm}
\usepackage{dsfont}
\usepackage{overpic}
\usepackage[dvipsnames]{xcolor}{\huge }
\usepackage[section]{placeins}
\definecolor{darkblue}{rgb}{0,0,.65}
\definecolor{darkgreen}{rgb}{0.28,0.41,0.19}
 
\usepackage{multirow}
\usepackage{tikz}
\usepackage{nicefrac}
\usepackage[
    pdfauthor={Nils Niggemann},
  pdfstartview=FitH,
  breaklinks=true,
  bookmarks=true,
  colorlinks=true,
  anchorcolor=black,
  citecolor=blue,
  filecolor=black,
  menucolor=black,
  urlcolor=darkblue,
  linkcolor=blue,
 ]{hyperref}

\usepackage[capitalize]{cleveref}

\newcommand{\e}{\mathrm{e}}
\definecolor{grey}{RGB}{180, 180, 180}

\usepackage[export]{adjustbox}
\newcommand{\subla}[1]{
{\includegraphics[width =#1\columnwidth]{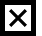}
}}

\newcommand{\sublX}{\mathchoice
  {\subla{0.03}}
  {\subla{0.03}}
  {\subla{0.018}}
  {\subla{0.015}}
}

\newcommand{\sublb}[1]{
{\includegraphics[width =#1\columnwidth]{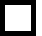}
}}

\newcommand{\sublO}{\mathchoice
  {\sublb{0.03}}
  {\sublb{0.03}}
  {\sublb{0.018}}
  {\sublb{0.015}}
}

\newcommand{\diagsquareXbare}[1]{{\includegraphics[width =#1\columnwidth]{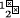}}}
\newcommand{\diagsquareObare}[1]{{\includegraphics[width =#1\columnwidth]{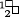}}}

\newcommand{\diagsquareO}{\mathchoice
  {\diagsquareObare{0.07}}
  {\diagsquareObare{0.05}}
  {\diagsquareObare{0.04}}
  {\diagsquareObare{0.04}}
}

\newcommand{\diagsquareX}{\mathchoice
  {\diagsquareXbare{0.07}}
  {\diagsquareXbare{0.05}}
  {\diagsquareXbare{0.04}}
  {\diagsquareXbare{0.04}}
}

\newcommand{\p}{\ensuremath\mathcal{P}}

\graphicspath{{images/}}

\begin{document}

\title{
Gapless fracton quantum spin liquid and emergent photons in a 2D spin-1 model}

\author{Nils Niggemann}
\email{nilsf.niggemann@gmail.com}
\affiliation{The Abdus Salam International Center for Theoretical Physics (ICTP), Strada
Costiera 11, I-34151 Trieste, Italy}
\affiliation{Helmholtz-Zentrum Berlin f\"ur Materialien und Energie, Hahn-Meitner Platz 1, 14109 Berlin, Germany}
\affiliation{Dahlem Center for Complex Quantum Systems and Fachbereich Physik, Freie Universit\"at Berlin, 14195 Berlin, Germany}

\author{Meghadeepa Adhikary}
\affiliation{Department of Physics and Quantum Centers in Diamond and Emerging Materials (QuCenDiEM) group, Indian Institute of Technology Madras, Chennai 600036, India}
\affiliation{SISSA, Via Bonomea 265, I-34136 Trieste, Italy}

\author{Yannik Schaden-Thillmann}
\affiliation{Helmholtz-Zentrum Berlin f\"ur Materialien und Energie, Hahn-Meitner Platz 1, 14109 Berlin, Germany}
\affiliation{Dahlem Center for Complex Quantum Systems and Fachbereich Physik, Freie Universit\"at Berlin, 14195 Berlin, Germany}

\author{Johannes Reuther}
\affiliation{Helmholtz-Zentrum Berlin f\"ur Materialien und Energie, Hahn-Meitner Platz 1, 14109 Berlin, Germany}
\affiliation{Dahlem Center for Complex Quantum Systems and Fachbereich Physik, Freie Universit\"at Berlin, 14195 Berlin, Germany}

\date{\today}

\begin{abstract}
Gapless fracton quantum spin liquids are exotic phases of matter described by higher-rank U(1) gauge theories which host gapped and immobile fracton matter excitations as well as gapless photons. Despite well-known field theories, no spin models beyond purely classical systems have been identified to realize these phases. Using error-controlled Green function Monte Carlo, here we investigate a square lattice spin-1 model that shows precise signatures of a fracton quantum spin liquid without indications of conventional ordering. Specifically, the magnetic response exhibits characteristic patterns of suppressed pinch points that accurately match the prediction of a rank-2 U(1) field theory and reveals the existence of emergent photon excitations in 2+1 spacetime dimensions. Remarkably, this type of fracton quantum spin liquid is not only identified in the system's ground state but also in generic low-energy sectors of a strongly fragmented Hilbert space.
\end{abstract}
\maketitle

\section{Introduction}
Quantum spin liquids (QSLs) are long-range entangled quantum phases with fractional spin excitations that cannot be smoothly deformed into conventional ordered phases~\cite{Balents2010,Savary-2017}. While elusive in nature, their theoretical description follows the established framework of gauge theories, in which different QSLs are characterized by different types of gauge fields~\cite{Wen2002}. In the most well-studied cases, gauge fields are of $\mathds{Z}_2$~\cite{Wegner1971} or U(1) type~\cite{Rantner2001} and are supplemented by `matter' fields, also referred to as spinons or charges $\rho$. Specifically, the U(1) case offers the remarkable possibility of realizing an emergent quantum electrodynamics (QED) theory in a QSL which manifests in an effective Gauss' law ${\bm \nabla}\cdot {\bm E}=\rho$ for the gauge field ${\bm E}$ as well as photon excitations~\cite{Hermele2004,Huse2003,benton2012}. Recently, it has been recognized that U(1) gauge fields allow for an intriguing generalization beyond conventional QED to higher-rank U(1) gauge theories~\cite{rasmussen2016stable,Pretko2017,Pretko2017_1,Xu-2006}, where the gauge fields $E^{\mu\nu}$ are matrices (or higher rank tensors). Specifically, in the rank-2 case and for scalar charges $\rho$ the generalized Gauss' law reads as $\partial_\mu\partial_\nu E^{\mu\nu}=\rho$. This modification gives rise to unconventional conservation laws, where not only $\rho$ but also the dipole moment ${\bm r}\rho$ is conserved which implies that charges $\rho$ -- in this case called fractons~\cite{Nandkishore-2019,Gromov2024,Pretko-2020,You2024} -- lose their mobility~\cite{Vijay-2015,Vijay-2016,Chamon-2005,Haah-2011}. In so-called type-I fracton theories~\cite{Chamon-2005,Bravyi-2011,Vijay-2015} dipoles of charges (named `lineons') still retain a partial mobility along subdimensional manifolds while in type-II fracton phases~\cite{Haah-2011,Yoshida-2013,Castelnovo-2012} all composite charges are immobile. In addition to matter particles a fractonic rank-2 U(1) QSL hosts gauge excitations given by gapless photons, similar to conventional QED. These gapless U(1) fracton systems should be contrasted with their {\it gapped} cousins, most prominently the X-Cube model~\cite{Vijay-2016,slagleFractonTopologicalOrder2017} and Haah's code~\cite{Haah-2011}, which exhibit $\mathds{Z}_2$-valued gauge fields giving rise to {\it fracton topological order}.

A general obstacle in the gauge theory description of QSLs is the difficulty of establishing a rigorous connection to microscopic interacting spin models. The knowledge of a parent spin Hamiltonian, however, is essential for the search and identification of QSLs in materials or synthetic platforms. Such a connection is only known for very few models. For example, the celebrated Kitaev honeycomb model~\cite{Kitaev-2006} exactly realizes a QSL where Majorana fermions couple to $\mathds{Z}_2$ gauge fields. Equally iconic, quantum spin ice~\cite{Gingras2014,Pace2021} as realized in a spin-1/2 XXZ model on the pyrochlore lattice can be mapped onto a compact U(1) gauge theory and, hence, gives rise to emergent QED~\cite{Hermele2004,Huse2003,benton2012}. In both cases, remarkable experimental progress has been made in recent years to identify these phases in magnetic materials~\cite{Singh2012,Banerjee2016,Banerjee2017,Banerjee2018,Pan2016,Sibille2018,Gaudet2019,Gao2019,Bhardwaj2022,Smith2022,Gao2024}. In contrast, higher-rank U(1) gauge theories for fractonic QSLs have so far only been identified in classical spin models on the purely electrostatic level~\cite{Yan2020,Benton2021,Davier2023,Yan2023_1,Yan2023_2,niggemannQuantumEffectsUnconventional2023}. The actual \emph{quantum} phase of a gapless fracton spin liquid has remained elusive~\cite{Sangeun2022} and its photon mode has completely resisted any description in terms of spin models. The difficulties in capturing fractonic properties in realistic spin systems also exist for gapped fracton models which typically require rather unrealistic multi-spin interactions~\cite{Vijay-2016,Haah-2011}.

\begin{figure*}
    \centering
    \includegraphics[width =\linewidth]{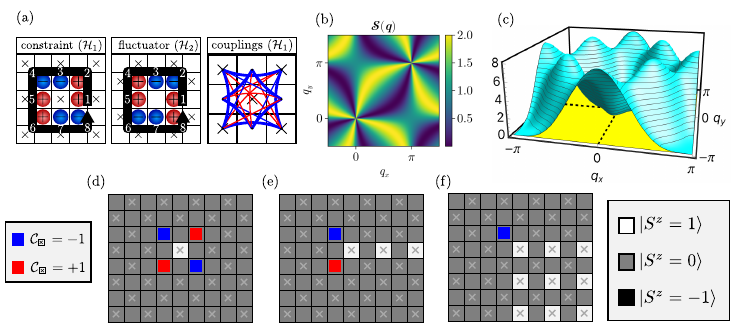}
    \caption{(a) Sign structure of the ground state constraints and fluctuators on eight-site clusters around sublattice 1 and sublattice 2 sites, respectively. The right panel shows the couplings that follow from squaring the constraint. Blue (red) couplings correspond to $J_{ij} = -1$ ($J_{ij} = 1$) and thick lines indicate interactions with $|J_{ij}| = 2$. (b) Spin structure factor in Gaussian approximation featuring fourfold pinch points. (c) Band structure of the coupling matrix $J_{ij}$ of $\mathcal{H}_1$ for $J=1$ with a dispersive upper band, flat lower band and quartic band touching points at $\bm q =(0,0)$ and $\bm q =(\pi,\pi)$. 
    (d) Acting with $S^+_i$ on a defect-free vacuum state $\ket{0,\dots,0}$ creates a quadrupole of four fractons with positive (red squares) and negative (blue squares) charges. (e) A \emph{lineon}, a dipole of fractons, which may be moved along the $x$-axis by single spin flips on the $\sublX$ sites. (f) Isolated fracton at the corner of the boundary between two distinct domains.}
    \label{fig:Overview}
\end{figure*}

In this paper, we fill this gap by introducing a surprisingly simple spin-1 model on the square lattice that imposes a rank-2 Gauss law constraint, leading to fractionalized excitations (fractons), a hallmark signature of QSLs. Most importantly, the magnetic response, calculated with error-controlled numerical methods, shows precise signatures of a rank-2 U(1) gauge theory, providing compelling evidence of a fracton QSL. Our results also reveal the distinct fingerprints of gapless gauge excitations, manifested in power-law spin correlations. Remarkably, this constitutes a realization of photons in 2+1 spacetime dimensions, previously thought impossible due to instanton proliferation~\cite{Polyakov1977}. As is typical for fracton systems, our model shows strong Hilbert space fragmentation, giving rise to macroscopically many dynamically disconnected subspaces even in the fracton-free sector of the Hilbert space, see also Supplementary Material for details. While such effects are detrimental to fractonic quantum properties in the spin-1/2 version of our model~\cite{Niggemann2025a}, we find that in the spin-1 case Hilbert space fragmentation can be helpful in this respect. Particularly, in addition to a fracton QSL in the system's ground state sector, we identify the same phase in generic excited sectors where its region of stability as a function of model parameters is even increased. This robustness of an emergent rank-2 U(1) gauge theory across different energy scales of our model may be helpful for experimental realizations of our theoretical predictions, e.g. on the basis of synthetic Rydberg atom platforms~\cite{browaeysManybodyPhysicsIndividually2020,giudiciDynamicalPreparationQuantum2022,Labuhn2016}.

\section{Model}
We define our model on the square lattice where we use the convention that the spins reside at the centers of the squares. For the definition of our model, we distinguish between the two sublattices of the square lattice, one marked with a cross ($\sublX$, sublattice 1) and the other drawn as empty squares ($\sublO$, sublattice 2) as shown in Fig.~\ref{fig:Overview}. The Hamiltonian consists of three terms $\mathcal{H}=\mathcal{H}_1+\mathcal{H}_2+\mathcal{H}_3$ given by
\begin{align}
\mathcal{H}_1&=\frac{J}{2}\sum_{\sublX}\mathcal{C}_{\sublX}^2,\notag\\ \mathcal{H}_2&=-J'\sum_{\sublO}\left(\mathcal{F}_{\sublO}+\mathcal{F}^\dagger_{\sublO}\right),\notag\\
\mathcal{H}_3&=\mu\sum_{\sublO}\left(\mathcal{F}^\dagger_{\sublO}\mathcal{F}_{\sublO}+\mathcal{F}_{\sublO}\mathcal{F}^\dagger_{\sublO}\right) \label{eq:Heff}
\end{align}
with
\begin{align}\label{eq:definition_cf}
\mathcal{C}_{\sublX}&=S^z_{\sublX_1}+S^z_{\sublX_2}-S^z_{\sublX_3}-S^z_{\sublX_4}+S^z_{\sublX_5}+S^z_{\sublX_6}-S^z_{\sublX_7}-S^z_{\sublX_8},\notag\\
\mathcal{F}_{\sublO}&=S^+_{\sublO_1}S^-_{\sublO_2}S^-_{\sublO_3}S^+_{\sublO_4}S^+_{\sublO_5}S^-_{\sublO_6}S^-_{\sublO_7}S^+_{\sublO_8}.
\end{align}
In this paper, $S_i^z$, $S_i^+$ and $S_i^-$ are spin-1 operators, and we refer to our companion paper~\cite{Niggemann2025a} for an investigation of the spin-$1/2$ model. Note the site labeling convention where $\sublO_a$ ($\sublX_a$) with $a=1,\ldots,8$ stands for one of the eight sites adjacent to $\sublO$ ($\sublX$), along horizontal, vertical and diagonal directions. Specifically, as illustrated in Fig.~\ref{fig:Overview}(a), the site $\sublO_1$ is located to the right of $\sublO$ and $\sublO_2$, $\sublO_3$, $\ldots$ progress counterclockwise around $\sublO$ (and the same for the sites $\sublX_a$). If no index $a$ is provided, as e.g. for $\mathcal{C}_{\sublX}$, the quantity is located directly at site $\sublX$. Furthermore, $i$ denotes a general site index not specifying the sublattice.

The sign structure $++--++--$ of the spin sums in $\mathcal{C}_{\sublX}$ and of the raising/lowering operators in $\mathcal{F}_{\sublO}$ are depicted in Fig.~\ref{fig:Overview}(a) and have close similarities. We call our model the {\it spiderweb model} since factoring out the squares in $\mathcal{C}_{\sublX}^2$ gives rise to a network of spin interactions resembling a spiderweb, see Fig.~\ref{fig:Overview}(a).

The general structure of $\mathcal{H}$ consisting of three terms is common to many spin liquid or quantum dimer models~\cite{Hermele2004,benton2012,Moessner2003,Sikora2011,Fancelli2024}. The first term $\mathcal{H}_1$ defines a low-energy subspace spanned by states that fulfill the constraints $\mathcal{C}_{\sublX}=0$ for the eight-site clusters around all $\sublX$, the so-called {\it constrained subspace}. The number of these states scales exponentially with the total number of sites (which also applies to the corresponding spin-1/2 model~\cite{Niggemann2025a}), see Supplementary Material for details. The second term $\mathcal{H}_2$ induces tunneling between states in the constrained subspace. Importantly, the so-called {\it fluctuators} $\mathcal{F}_{\sublO}$ commute with each of the constraint operators $\mathcal{C}_{\sublX}$, such that the quantum dynamics generated by $\mathcal{H}_2$ does not lead the system out of the constrained subspace.

While $\mathcal{H}_1$ consists of usual two-body spin interactions, the eight-site ring-exchange term $\mathcal{F}_{\sublO}$ might first seem complicated and artificial. However, this term represents the shortest product of spin flip operators that generates tunneling between states within the constrained subspace. This implies that {\it any} small perturbation of $\mathcal{H}_1$ that is off-diagonal in the Ising $S_i^z$ basis will inevitably generate $\mathcal{H}_2$ when projected onto the constrained subspace. For example, $\mathcal{H}_2$ is generated in fourth order perturbation theory for small transverse nearest neighbor couplings $S_i^x S_j^x+S_i^y S_j^y$ or in eighth order perturbation theory in small transverse magnetic fields $\sim S_i^x$. These properties are similar to quantum spin ice where minimal hexagon loop moves are generated in third order perturbation theory in transverse nearest neighbor couplings~\cite{Hermele2004,benton2012}.

The third term $\mathcal{H}_3$ counts the number of eight-site clusters that are not annihilated by $\mathcal{F}_{\sublO}$ or $\mathcal{F}^{\dagger}_{\sublO}$. It can be understood as a chemical potential for flippable clusters and has been introduced in many other spin liquid models before, such as short-range $\mathds{Z} _2$~\cite{Moessner2001,Moessner2003,Fancelli2024} or algebraic pyrochlore U(1) spin liquids~\cite{Moessner2003,Hermele2004,benton2012,Sikora2011}. By changing $\mu$ the system can be tuned through different phases. Specifically, the limits $\mu\rightarrow \infty$ ($\mu\rightarrow -\infty$) where the numbers of flippable clusters are minimized (maximized) typically give rise to simple ordered states. On the other hand, at $\mu\in(0,J']$ the competition between kinetic ($\mathcal{H}_2$) and potential ($\mathcal{H}_3$) energy may create non-trivial strongly fluctuating quantum phases. The system with $\mu=J'>0$, known as the Rokhsar-Kivelson point~\cite{Rokhsar1988}, is exactly solvable where the ground state is an equal weight superposition of all states in the constrained subspace. More precisely, if the constrained subspace is again divided into dynamically disconnected sectors by the action of $\mathcal{F}_{\sublO}$ and $\mathcal{F}_{\sublO}^{\dagger}$ (as is the case in our system), a ground state can be constructed from the equal weight superposition of states in {\it each} of these dynamically disconnected sectors. 

For large enough sectors, such a massive superposition can be associated with a quantum spin liquid. However, whether this spin liquid also exists as an extended phase for $\mu < J'$ is highly non-trivial and model dependent.

\section{General fractonic properties}
Our comprehensive numerical investigations presented below show that the spin-1 spiderweb model exhibits a fracton QSL in a finite region in $\mu$. These studies can be understood as a continuation of our investigations of the corresponding spin-1/2 model in our companion paper~\cite{Niggemann2025a}. However, that work revealed that the quantum dynamics induced by $\mathcal{H}_2$ are too weak to generate a QSL. The reason for the suppressed quantum dynamics in the spin-1/2 case is a severe Hilbert space fragmentation as is typical for fracton models~\cite{Adler2024,Will2024,Feng2022}. Specifically, $\mathcal{H}_2$ splits up the constrained subspace into an even finer structure of many dynamically disconnected subspaces, each with only trivial quantum dynamics. This results in either long-range ordered phases or {\it classical fracton spin liquids}, i.e. ground states characterized by a classical rank-2 gauge theory without coherent quantum dynamics. As we will see below, the availability of more spin states $S_i^z\in\{-1,0,1\}$ for spin-1 significantly enhances the quantum dynamics and promotes fractonic properties at a quantum level. Note however that Hilbert space fragmentation is still present in the spin-1 model and again gives rise to a vast number of dynamically isolated subspaces scaling exponentially in the number of sites, as discussed in the Supplementary Material.
 
While the identification of a fracton QSL in Sec.~\ref{sec:spin_one} requires advanced numerical approaches, some general fractonic properties of the spiderweb model are already evident from its structure and are independent of the spin magnitude. 
Here, we explain the key properties and refer to our companion paper~\cite{Niggemann2025a} for a more in-depth discussion.

\subsection{Gaussian approximation}
The Gaussian approach~\cite{Yan2023_1,Yan2023_2,Benton2021} consists of treating the spins as unconstrained variables (no local normalization or quantization imposed) and in Fourier space $S^z_m({\bm q})=\sum_{i\in m}e^{\imath {\bm q}\cdot {\bm r}_i} S_i^z$  where $m=1,2$ denotes the two sublattices. Expanded in lowest non-vanishing order (quadratic order) around ${\bm q}=0$ the classical constraints $\mathcal{C}_{\sublX}=0$ then take the form
\begin{equation}\label{eq:gauss_law_s}
(q_x^2-q_y^2)S_2^z({\bm q})+4q_x q_y S_1^z({\bm q})=0\;.
\end{equation}
Arranging the Fourier-components in a symmetric and trace-free matrix $\underline{S}({\bm q})$ defined as
\begin{equation}
\underline{S}({\bm q})=\left(\begin{array}{cc}
S_2^z({\bm q}) & 2S_1^z({\bm q})\\
2S_1^z({\bm q}) & -S_2^z({\bm q})\end{array}\right)
\end{equation}
the constraint can be written compactly as $q_\mu q_\nu \underline{S}^{\mu\nu}({\bm q})=0$ which is exactly the momentum space version of the generalized charge-free Gauss' law of a trace-free rank-2 U(1) electrostatic theory, $\partial_\mu\partial_\nu E^{\mu\nu}=0$, with a fictitious matrix-valued `electric' field $E^{\mu\nu}$~\cite{Pretko2017,Pretko2017_1,Prem2018,Nandkishore-2019,Pretko-2020} (see also the Supplementary Material for details on the derivation of the Gauss law). Violations of the constraint $\partial_\mu \partial_\nu E^{\mu\nu}=\rho\neq0$ take the role of charges which in this higher-rank case are called (scalar type-I) fractons with no mobility in real space. We note that an identical rank-2 Gauss' law also follows from an expansion of the constraint around ${\bm q}=(\pi,\pi)$ indicating that long-wavelength fluctuations around ferromagnetic and antiferromagnetic states are both of fractonic nature. As explained in Ref.~\cite{Niggemann2025a} the emergent rank-2 Gauss' laws are a direct consequence of the special sign structure of the spin sum in $\mathcal{C}_{\sublX}$ corresponding to discretized second derivatives.

The rank-2 Gauss' law manifests in singular points in the spin structure factor 
\begin{equation}
    \mathcal{S}({\bm q})=\frac{1}{N_{\textrm{sites}}}\sum_{mn}\langle S^z_m(-{\bm q})S^z_n({\bm q})\rangle \label{eq:StructureFac}
\end{equation}
 at $T=0$ known as fourfold pinch points~\cite{Prem2018}. Within the present Gaussian formulation these features are obtained by carrying out $\langle\cdots\rangle$ as a projection onto the Fourier-modes that fulfill Eq.~(\ref{eq:gauss_law_s}). The result shown in Fig.~\ref{fig:Overview}(b) features clear fourfold pinch points in the extended Brillouin zone $q_x,q_y \in [0,2\pi)$ at ${\bm q}=(0,0)$ and ${\bm q}=(\pi,\pi)$ which are just the points where an expansion of the constraint has the form of a rank-2 Gauss' law.

An alternative description of the same properties is obtained by writing $\mathcal{H}_1$ as a $2\times 2$ matrix in the Fourier-components $S_1^z({\bm q})$ and $S_2^z({\bm q})$. As shown in Fig.~\ref{fig:Overview}(c) the diagonalization of this matrix results in a flat bottom band (describing the constrained subspace) which is connected to an upper dispersive band at two band touching points which occur at the exact momenta of the pinch points. The fourfold nature of pinch points implies that the dispersive band is {\it quartic} in ${\bm q}$ in the vicinity of the band touching points.

\subsection{Classical fracton configurations from \texorpdfstring{$\mathcal{H}_1$}{H1}}
A particularly intuitive access to the fractonic properties of the spiderweb model is obtained by examining classical fracton configurations in real space as shown in Fig.~\ref{fig:Overview}(d), (e) and (f). In Fig.~\ref{fig:Overview}(d), a single $S_i^z=1$ site on sublattice 1 in a homogeneous $S_i^z=0$ background (which is one possible ground state of $\mathcal{H}_1$) corresponds to four violated constraints $\mathcal{C}_{\sublX}\neq 0$ (fractons) forming a quadrupole. This implies that a single spin flip fractionalizes into four fractons, directly demonstrating that fractionalization -- a hallmark of a QSL -- is already intrinsic to $\mathcal{H}_1$. These four violated constraints constitute the smallest group of fractons that can move freely as a whole by lowering the changed spin back to zero and raising a neighboring one. On the other hand, a semi-infinite string of $S_i^z=1$ sublattice 1 sites [Fig.~\ref{fig:Overview}(e)] gives rise to a dipole of fractons (so-called lineons~\cite{Pai2019}) which can move perpendicular but not parallel to its dipole moment. Finally, a corner in the domain wall of Fig.~\ref{fig:Overview}(f) corresponds to an isolated fracton which is completely immobile, unless infinitely many spins are flipped (moving the lower right domain as a whole) or additional fractons are created. This immobility is a direct consequence of the dipol conservation which is again a result of the emergent rank-2 Gauss' law~\cite{Nandkishore-2019}.

\subsection{Conserved magnetizations}
\begin{figure}
    \centering \includegraphics[width=1\linewidth]{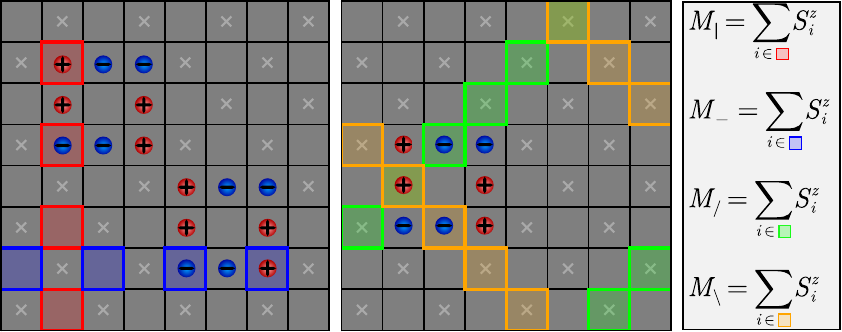}
    \caption{String-like magnetizations $M_{\bm \vert}$, $M_{\bm -}$, $M_{\bm \diagdown}$, $M_{\bm \diagup}$ which are conserved under the application of the fluctuator $\mathcal{F}_{\sublO}$. Colored squares correspond to the sites which are summed over. Also shown is the action of $\mathcal{F}_{\sublO}$ at exemplary locations which always cause canceling contributions to any of the operators $M_{\bm \vert}$, $M_{\bm -}$, $M_{\bm \diagdown}$, and $M_{\bm \diagup}$. }
    \label{fig:conservation_laws}
\end{figure}

Typical fracton properties are also reflected in the constants of motion of our spiderweb model. From the definition of $\mathcal{F}_{\sublO}$ in \cref{eq:definition_cf}, also shown in \cref{fig:Overview}(a), it can be seen that the total magnetization $M^z_{m} = \sum_{i \in m} S_i^z$ in each sublattice $m=1,2$ and the (spin) dipole moment $\sim \sum_i \bm{r}_i S^z_i$ are conserved. In addition to these global conserved magnetizations the system also has subdimensional constants of motion as they are characteristic for type-I fracton models. In our case they correspond to spin sums $M_{\bm \diagdown},\ M_{\bm \diagup} ~ (M_{\bm \vert},\ M_{\bm -})$ on diagonal (straight) lines containing only sites of sublattice $1$ (sublattice $2$), see Fig.~\ref{fig:conservation_laws}. These subdimensional conserved magnetizations can be viewed in analogy to the subsystem symmetries known from \emph{gapped} fracton models~\cite{Xu2004,Vijay-2016,yanHyperbolicFractonModel2019a,ibieta-jimenezFractonlikePhasesSubsystem2020}.

\section{Results}\label{sec:spin_one}
\begin{figure*}[t]
    \centering
    \includegraphics[width=0.99\linewidth]{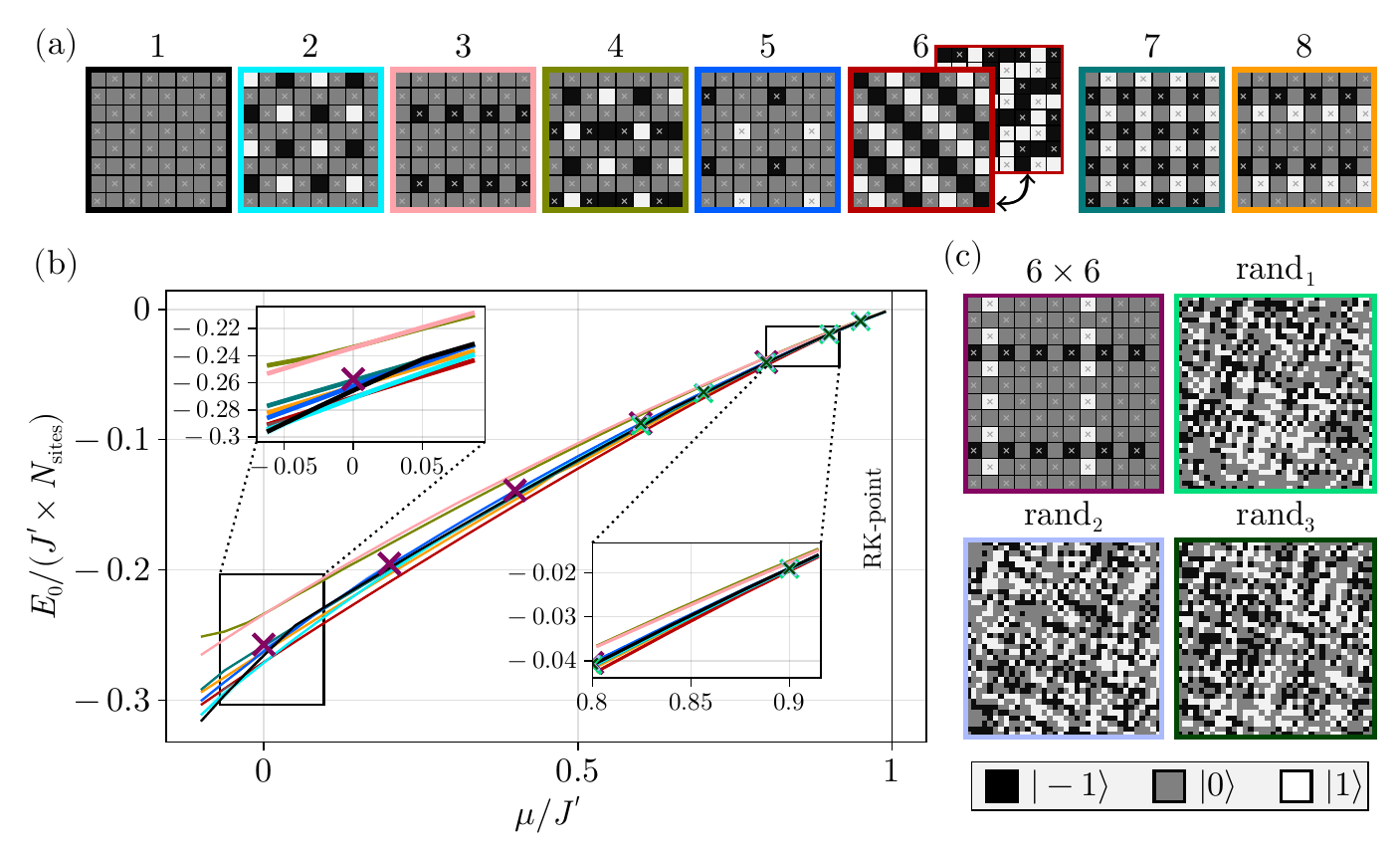}
    \caption{(a) Maximally flippable configurations defining the eight most energetically favorable sectors with a $4\times4$ unit cell, so-called parent states. (b) Energy of each sector shown in (a) (solid lines) and (c) (markers) as a function of $\mu$. (c) Other configurations that define individual sectors such as a configuration with a $6\times6$ unit cell as well as randomized configurations. For the configurations in (a), systems sizes of $L=20$ are used while for (c) the system sizes are $L=24$ for the $6\times6$ sector and $L=36$ for the random sectors.}
    \label{fig:SectorEnergies}
\end{figure*}
\subsection{Ground state properties}\label{sec:ground_state}
We shall now discuss the ground state quantum properties of \cref{eq:Heff}, in the limit $J' \ll J$, where we need to only consider fracton-free sectors for which $\mathcal{H}_1 = 0$.
For $J'>0$ our model is free of the sign problem and thus observables may be obtained numerically within statistical error bars via quantum Monte Carlo (QMC). As we are interested in ground state properties, we choose the so-called Green function Monte Carlo (GFMC) method which allows sampling of ground state observables within a given ergodicity sector of $\mathcal{H}$ \cite{Buonaura1998,becca2017quantum}. Details regarding this method are summarized in \cref{app:gfmc}.

To identify the sector containing the ground state out of an exponentially large number of sectors, we first generate {\it all} periodic fracton-free spin configurations (in $S^z$ basis) with a $4\times4$ unit cell. Our numerical simulations are then performed on larger $L\times L$ systems (with $L$ up to 36) where the starting configurations correspond to periodic tilings of the $L\times L$ system with these $4\times 4$ unit cells. Determining the ground states in the sectors connected to each of these $4\times4$ `parent' states we systematically scan large numbers of subspaces. Specifically, discarding configurations with no flippable clusters and eliminating redundant configurations using point-group or time reversal symmetries, we find 1104 periodic $4\times4$ configurations. Under the application of $\mathcal{H}_2$ these configurations give rise to 28 fully inequivalent, disconnected Hilbert space sectors, which can be distinguished via their energetic properties and their \emph{maximally flippable configuration}, which maximizes $\sum_{\sublO} \mathcal{F}^\dagger_{\sublO}\mathcal{F}_{\sublO}+ \mathcal{F}_{\sublO}\mathcal{F}^\dagger_{\sublO}$.
These configurations are depicted for the eight lowest energetic sectors in \cref{fig:SectorEnergies}(a).
Clearly, this approach does not capture all Hilbert space sectors. For example, it is in principle possible that the ground state lies in another sector connected to periodic $6\times 6$ tilings (however, an exhaustive enumeration of such tilings proved to be numerically infeasible due to the large number of solutions). In that case, however, already simulations of sectors from periodic $4\times 4$ states are expected to show traces of 6-site periodic correlations such as Bragg peaks in the spin structure factor at momenta ${\bm q}=2\pi(1/6,1/6)$. This particularly applies to the regime $\mu\ll J'$ where systems tend to establish long-range order built from periodic tilings. The absence of such Bragg peaks in our results for $4\times 4$ parent states indicates that our approach includes enough sectors to identify the overall ground state.

In \cref{fig:SectorEnergies}(b), we show the energy of the eight energetically lowest-lying sectors as a function of $\mu$, where \cref{fig:SectorEnergies}(a) illustrates the corresponding eight $4\times 4$ periodic parent states. Figure \ref{fig:SectorEnergies}(b) also displays the energies of other sectors shown in \cref{fig:SectorEnergies}(c), such as one with a larger $6\times 6$ unit cell as well as randomly sampled fracton-free configurations without any periodic parent states. 

The most interesting regime is at $0<\mu\leq J'$ where the competition between kinetic and potential energy increases quantum fluctuations. Here, we find that the system's ground state lies in the sector of the diagonal stripe state [shown as sector number 6 (foreground) in \cref{fig:SectorEnergies}(a)]. Interestingly, this sector also contains the state shown in the background of sector number 6 in \cref{fig:SectorEnergies}(a) which has a ``staircase-pattern'' built from local spin states $S^z_i=\pm1$. This configuration is an analogue of the staircase state discussed in Ref.~\cite{Niggemann2025a} in the context of the spin-1/2 spiderweb model, where the ground state consists of fluctuations around a staircase configuration with $S^z_i=\pm1/2$.

Surprisingly, the sector of the homogeneous $S_i^z=0$ state [number 1 in \cref{fig:SectorEnergies}(a)] is an excited sector across the entire regime $0\leq\mu< J'$ even though it allows for substantial quantum fluctuations: This state has the unique property that \emph{all} clusters $\sublO$ are flippable by $\mathcal{F}_{\sublO}$ \emph{and by} $\mathcal{F}^\dagger_{\sublO}$ and thus exhibits the absolute maximum number of flippable clusters among all spin-1 Ising configurations.
With this property it is clear that this sector must become the ground state sector for sufficiently small $\mu$. Our results in \cref{fig:SectorEnergies}(b) confirm this to be the case for $\mu\lesssim0$. With decreasing $\mu$ the negative potential term suppresses fluctuations into local $S_i^z=\pm1$ configurations and in the limit $\mu\rightarrow -\infty$ the ground state continuously transforms into the exact homogeneous $S_i^z=0$ state. Thus, even though no symmetry breaking occurs in this regime and no long-range order is observed the system is a trivial paramagnet that is continuously connected to a product state ($S_i^z=0$ on every site).

In the other limit $\mu> J'$, $\mathcal{H}$ is positive semi-definite. There, the exact ground state is given by trivial sectors containing only a single configuration with no flippable clusters which is annihilated by $\mathcal{H}$. While there is still a substantial number of these sectors, it is clear that no quantum dynamics can occur in the ground state for $\mu> J'$ and the system behaves classically.
\begin{figure}
    \centering
    \includegraphics[width=\linewidth]{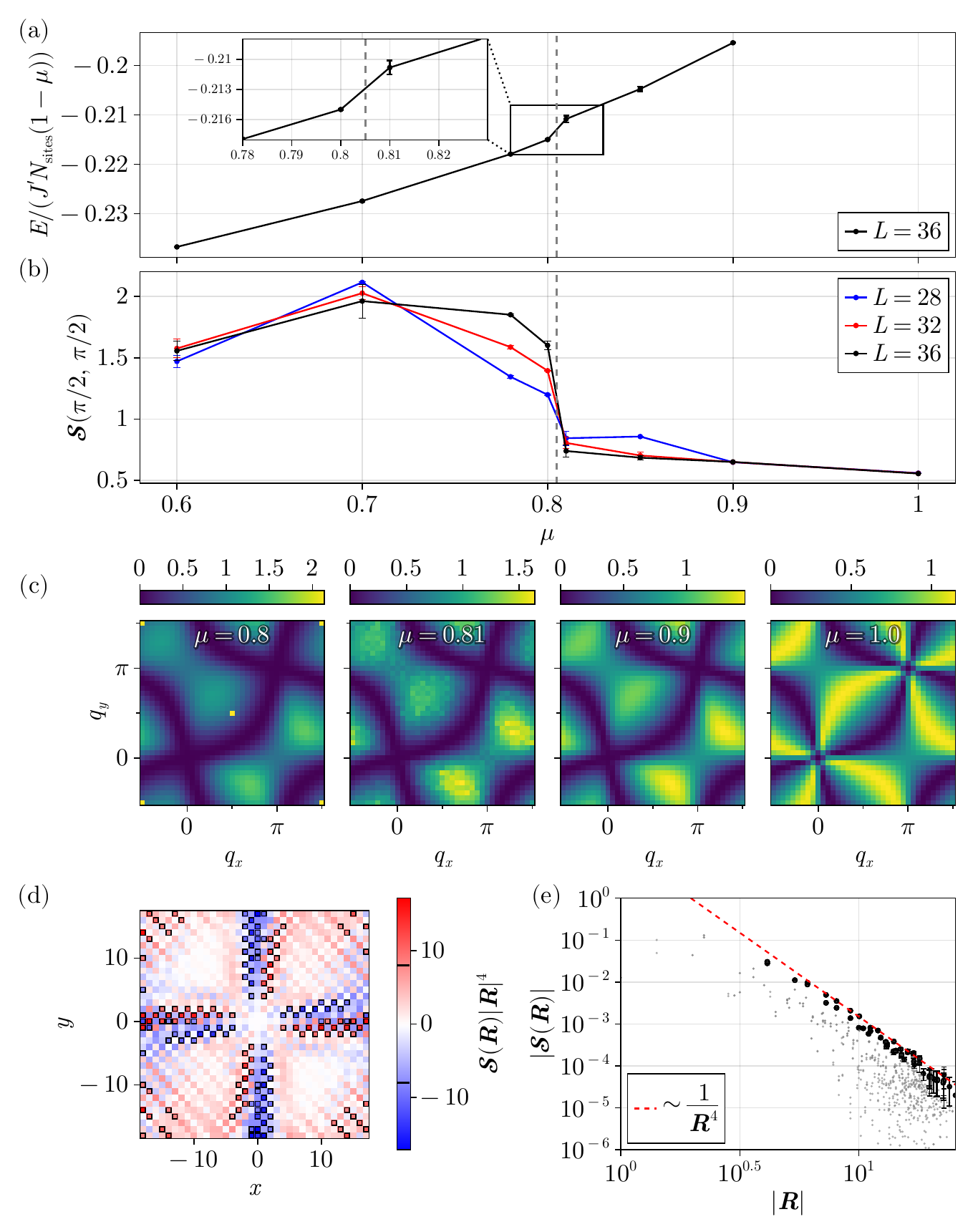}
    \caption{(a) Energy per site in the diagonal stripe sector scaled by $(J'-\mu)^{-1}$. The dashed line indicates the location of the energy's discontinuity. (b) Ground state spin structure factor [\cref{eq:StructureFac}] in the diagonal stripe sector from GFMC at ${\bm q} = (\pi/2,\pi/2)$ as a function of $\mu$ for different system sizes $L$. (c) Momentum resolved ground state spin structure factor in the diagonal stripe sector from GFMC for different $\mu$ and $L=36$. (d) Direct space spin correlations $\mathcal{S}({\bm R})$ for $\mu=0.9J'$, scaled by $|{\bm R}|^4$, with dominant correlations highlighted in black. (e) Decay of spin correlations as a function of distance $|{\bm R}|$ compared to a power-law $\sim |{\bm R}|^{-4}$ (red dashed line) for $\mu=0.9J'$. Grey (black) markers indicate generic (dominant) correlations. }
    \label{fig:Stair_Sq}
\end{figure}

Having identified the ground state sector at $0\leq\mu<J'$, it remains to be determined whether the long-range correlations in its diagonal stripe parent state can be wiped out by quantum fluctuations giving rise to a spin liquid. Even at the RK point $\mu = J'$ where the exact wave function is an equal weight superposition of all states in that sector, it is not a priori clear whether a QSL forms. Indeed, for spin $S=1/2$, it was found that the large degree of Hilbert space fragmentation competes with quantum dynamics and leads to strongly classical behavior with negligible impact of quantum fluctuations even at the RK-point~\cite{Niggemann2025a}. Beside this, spin liquids in two-dimensional quantum dimer models are often found to have a vanishing stability regime, such that the RK point behaves as a quantum critical point between two ordered phases, for example on bipartite lattices~\cite{Rokhsar1988,zengQuantumDimerModels2025,senthilQuantumCriticalityLandauGinzburgWilson2004,fradkinBipartiteRokhsarKivelsonPoints2004}.

In investigation of this, \cref{fig:Stair_Sq}(a) shows an enlarged view of the energy per site in the diagonal stripe sector as a function of $\mu$, rescaled by a factor of $(J'-\mu)^{-1}$ for better visibility of features beside the trivial energetic scaling $\sim -(J'-\mu)$. A kink in the energy around $\mu=0.8 J'$ indicates a first order phase transition. To distinguish these phases, \cref{fig:Stair_Sq}(b) shows the value of the spin structure factor at ${\bm q} = (\pi/2,\pi/2)$, the momentum of the diagonal stripe order as a function of $\mu$. Here, a much stronger discontinuity can be seen between $\mu=0.8J'$ and $\mu=0.81J'$ which becomes increasingly pronounced with the system size. Inspecting the full momentum dependence of $\mathcal{S}(\bm{q})$ in \cref{fig:Stair_Sq}(c), gives further insight into the nature of the two phases. For $\mu \leq 0.8J'$, the spin structure factor shows a clear peak at ${\bm q} = (\pi/2,\pi/2)$, signaling conventional magnetic long-range order. For $\mu >0.8J'$, this peak disappears abruptly, revealing fourfold pinch points that are suppressed around their respective centers located at ${\bm q} = (0,0)$ and ${\bm q} = (\pi,\pi)$. Indeed, this pinch point suppression is a well-known signature of emergent photons in conventional U(1) gauge theories and has been numerically confirmed for quantum spin ice~\cite{benton2012}. 
As $\mu$ increases, the intensity around the pinch point is continuously restored, and, finally, at the RK point precisely corresponds to the Gaussian approximation from \cref{fig:Overview}(b)
\footnote{Note that the $+$-shaped minima at the pinch point origin are finite-size effects which always correspond to the four momenta closest to the pinch point at \emph{any} system size $L$ with their volume shrinking to zero for $L \rightarrow\infty$.}. We emphasize that the appearance of a fourfold pinch point within a \emph{single sector} of the Hilbert space is in contrast to spin-$1/2$, where this feature only appears after summing classical contributions from many sectors~\cite{Niggemann2025a}. 

\Cref{fig:Stair_Sq}(d) displays the directional dependence of the real space correlation function $\mathcal{S}({\bm R}) = \frac{1}{N}\sum_i \expval{S^z(\bm{r}_i) S^z(\bm{R} - \bm{r_i})}$ at $\mu=0.9J'$. The rescaling by $|{\bm R}|^4$ highlights a checkerboard pattern of weak and strong correlations with dominant directions near the Cartesian axes. \Cref{fig:Stair_Sq}(e) shows the distance dependence of the correlation function, indicating a power-law $\sim |{\bm R}|^{-4}$ behavior consistent with a gapless state~\cite{Hastings2006} (see Supplementary Material for an explanation of the $|{\bm R}|^{-4}$ scaling).

These findings provide strong evidence of a gapless QSL at the RK point which remains stable in a finite region $0.81J' \leq \mu \leq J'$. Later, this result will be further substantiated using more rigorous field theory arguments.

A peculiar feature of \cref{fig:Stair_Sq}(c) is the rotational asymmetry around the pinch points. Specifically, at $\mu<J'$, the $90^\circ$ rotation symmetry of a perfect fourfold pinch point is lowered to a two-fold symmetry, giving rise to two of four lobes with relatively higher intensity. 
Towards the RK point $\mu=J'$, this fourfold rotational symmetry is continuously restored.
The rotational asymmetry at $\mu<J'$ is a consequence of the fact that the diagonal stripe configuration itself breaks $90^\circ$ rotation symmetry which is not restored by quantum fluctuations. In fact, it can be checked that a simulation starting from a $90^\circ$ rotated diagonal stripe state yields a $90^\circ$ rotated spin structure factor. This implies that the symmetry breaking is {\it not spontaneous} but rather imposed by the considered parent state. This latter property is further substantiated in the Supplementary Material where it is shown that $90^\circ$ rotated versions of the diagonal stripe configuration lie in independent, dynamically disconnected sectors. Therefore, the broken rotational symmetry is not contradictory to a QSL. In fact, in the next section we will show that it is fully compatible with a spin liquid obeying an emergent rank-2 U(1) gauge structure.

\subsection{Emergent rank-2 U(1) quantum electrodynamics}
The results above indicate that the spin-1 spiderweb model realizes a stable QSL phase for $0.8 J'<\mu \leq J'$. Characterized by quantum fluctuations within a manifold of configurations satisfying a rank-2 Gauss' law, this phase is thus expected to be a higher-rank U(1) QSL.
To further verify this expectation we turn to an effective field theoretical description of \cref{eq:Heff}. 
Indeed, the well-known mapping of quantum spin ice to a U(1) gauge theory~\cite{Huse2003,Hermele2004,benton2012} can be transferred to our higher-rank system. Following a similar strategy, we express the spin operators ${\bm S}_i$ in terms of conjugate rotor variables $A_i^{xy/xx}$ and $E_i^{xy/xx}$ via
\begin{align}
S_i^\pm&=\begin{cases}\sqrt{2}e^{\pm \imath A^{xy}_i}& i=\text{center of\ }\sublX \text{\ (sublattice 1)}\\
\sqrt{2}e^{\pm \imath A^{xx}_i}& i=\text{center of\ }\sublO \text{\ (sublattice 2)}
\end{cases},\label{eq:a_def}\\
S_i^z&=\begin{cases}E_i^{xy}&\quad\quad\  i=\text{center of\ }\sublX \text{\ (sublattice 1)}\\
E_i^{xx}&\quad \quad \ i=\text{center of\ }\sublO \text{\ (sublattice 2)}
\end{cases},\label{eq:e_def}
\end{align}
with the commutation relations $[A_i^{xy},E_j^{xy}]=[A_i^{xx},E_j^{xx}]=\imath\delta_{ij}$ and $[A_i^{xy},E_j^{xx}]=[A_i^{xx},E_j^{xy}]=0$. Note that the factor $\sqrt{2}$ in Eq.~(\ref{eq:a_def}) accounts for the spin-1 quantum number. As conjugate rotor variables, the fields $A_i^{xy/xx}\in[0,2\pi]$ are compact while the `number operators' $E_i^{xy/xx}\in\mathds{Z}$ are integer, and can thus be different from ${-1,0,1}$ in principle.
Furthermore, $A_i^{xy/xx}$ correspond to the components of a $2\times 2$ trace-free ($A_i^{xx}=-A_i^{yy}$) and symmetric ($A_i^{xy}=A_i^{yx}$) matrix-valued field that takes the role of a rank-2 generalization of the vector potential in conventional electrodynamics. Similarly, $E_i^{xy/xx}$ are the components of a matrix-valued electric field with the analogous properties $E_i^{xx}=-E_i^{yy}$ and $E_i^{xy}=E_i^{yx}$.
We may now define a gauge invariant emergent magnetic field $B_{\sublO}$ (see Supplementary Material for details) for each center of a $\sublO$ cluster via
\begin{equation}\label{eq:bfield}
B_{\sublO}=A^{xy}_{\sublO_1}-A^{xx}_{\sublO_2}-A^{xy}_{\sublO_3}+A^{xx}_{\sublO_4}+A^{xy}_{\sublO_5}-A^{xx}_{\sublO_6}-A^{xy}_{\sublO_7}+A^{xx}_{\sublO_8},
\end{equation}
where the sign pattern follows that of the fluctuator [Fig.~\ref{fig:Overview}(a)] and the notation for sites ${i=\sublO_a}$ is the same as in Eq.~(\ref{eq:Heff}).

With these ingredients we can now formulate an effective field theory for $\mathcal{H}$ in the constrained subspace,
\begin{align}
\mathcal{H}_{\text{eff}}&=\frac{U}{2}\left[\sum_{\sublX} (E_{\sublX}^{xy})^2+\sum_{\sublO} (E_{\sublO}^{xx})^2\right]\notag\\
&+\frac{K}{2}\sum_{\sublO}B^2_{\sublO}+\frac{W}{2}\sum_{\sublO} \mathcal{N}_{\sublO}^2,\label{eq:maxwell}
\end{align}
with
\begin{equation}\label{eq:nfield}
\mathcal{N}_{\sublO}=E^{xy}_{\sublO_1}-E^{xx}_{\sublO_2}-E^{xy}_{\sublO_3}+E^{xx}_{\sublO_4}+E^{xy}_{\sublO_5}-E^{xx}_{\sublO_6}-E^{xy}_{\sublO_7}+E^{xx}_{\sublO_8}.
\end{equation}
The first term $\sim U$ suppresses high values of $|E^{xy/xx}_i|$ to (approximately) enforce the constraint $S_i^z=-1,0,1$. This term describes the energy density of the electric field and has the same form as the corresponding term in a usual Maxwellian field theory.

The second term $\sim K$ comes from inserting Eq.~(\ref{eq:a_def}) into $\mathcal{H}_2$ and using Eq.~(\ref{eq:bfield}). This gives $\mathcal{H}_2\sim-\cos B_{\sublO}$ and, when expanded to quadratic order, yields $\sim B^2_{\sublO}$, corresponding to the energy density of the magnetic field, as in a Maxwellian field theory. The assumption behind the expansion that $B_{\sublO}$ fluctuates only mildly around $B_{\sublO}=0$, however, is not necessarily fulfilled. In fact, phase slip events between two minima of the cosine $B_{\sublO}\rightarrow B_{\sublO}+2\pi$ that cannot be captured within a finite-order expansion are a known phenomenon of U(1) field theories in 2+1 dimensional spacetime~\cite{Polyakov1977}. If these so-called \emph{instanton} events proliferate, they may drive the system into an ordered and confined phase. By comparing our numerical results for the spin-1 spiderweb model with the predictions of the effective field theory we will confirm that the assumptions and approximations behind Eq.~(\ref{eq:maxwell}) are justified.

The last term $\sim W$ in Eq.~(\ref{eq:maxwell}) mimics the potential term $\mathcal{H}_3$, where the exact RK point is realized in the limit $U/K\rightarrow0$~\cite{benton2012}. Following the construction of Refs.~\cite{Hermele2004,benton2012}, $\mathcal{N}_{\sublO}$ is obtained from $B_{\sublO}$ by replacing $A^{xy}\rightarrow E^{xy}$ and $A^{xx}\rightarrow E^{xx}$ in Eq.~(\ref{eq:bfield}).

The model in Eq.~(\ref{eq:maxwell}) is a free bosonic theory describing a rank-2 U(1) QSL. Importantly, it can be solved exactly, yielding a single photon mode $\omega(\bm q)$ with gapless nodal points at the pinch point locations, see \cref{fig:Condensate_mu_Sweep_xi}(d). For $U>0$, away from the RK point, an expansion of $\omega(\bm q)$ in $q$ around the pinch points yields in lowest orders
\begin{equation}\label{eq:photon}
\omega(\bm q)\approx\sqrt{\frac{KU}{4}}\sqrt{q_x^4+14 q_x^2 q_y^2+q_y^4}\;,
\end{equation}
revealing a \emph{quadratic} photon dispersion $\omega({\bm q})\sim q^2$ for small $q$ along any radial direction away from ${\bm q}=0$. Exactly at the RK point $U=0$ the photon dispersion becomes even quartic, $\omega({\bm q})\sim q^4$. The field theory also predicts the spin structure factor $\mathcal{S}({\bm q})$ which we can compare with the numerical results from GFMC by taking the parameters $U$, $K$ and $W$ as fit parameters [see Supplementary Material for an explicit analytical expression].

In principle, the field theory in Eq.~(\ref{eq:maxwell}) is applicable to any fracton-free Hilbert space sector of the spiderweb model. However, as described in Sec.~\ref{sec:ground_state} in the context of the diagonal stripe state, specific sectors may imprint a rotational asymmetry. To account for such effects, we add an additional phenomenological term to the Hamiltonian:
\begin{equation}\label{eq:asymmetric}
    \mathcal{H}_{\text{eff}} \rightarrow \mathcal{H}_{\text{eff}} + \frac{U'}{2}\left[ \sum_{\diagsquareX}(E^{xy}_{1} + E^{xy}_{2})^2 + \sum_{\diagsquareO} (E^{xx}_{1} + E^{xx}_{2})^2\right].
\end{equation}
It corresponds to a modification of the first term in Eq.~(\ref{eq:maxwell}) where the local $E_i^2$-terms are made non-local by coupling electric fields $E_1$, $E_2$ on pairs of sites separated along one diagonal direction, \includegraphics[scale=1]{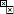} and \includegraphics[scale=1]{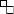}, therefore breaking $90^\circ$ rotation symmetry. Importantly, this gauge invariant term only affects the directional shape of the emergent photon dispersion and spin structure factor at {\it small} wavelengths. The long wavele ngth limit remains unaffected retaining the exact gapless and $90^\circ$ rotation-symmetric photon dispersion of Eq.~(\ref{eq:photon}).

To obtain the best fit to these field theory parameters, we define three independent fitting parameters $(A,r,p)$ through $(K,W,U,U') = (4A^2,1,r,pr)$. In the spin structure factor, $A$ provides a trivial uniform scaling, while $r$ interpolates between the RK point at $r=0$ and the RK-free limit $r = \infty$, and $p = U'/U$ is the strength of the rotational asymmetry.

\begin{figure}
    \centering
    \includegraphics[width=\linewidth]{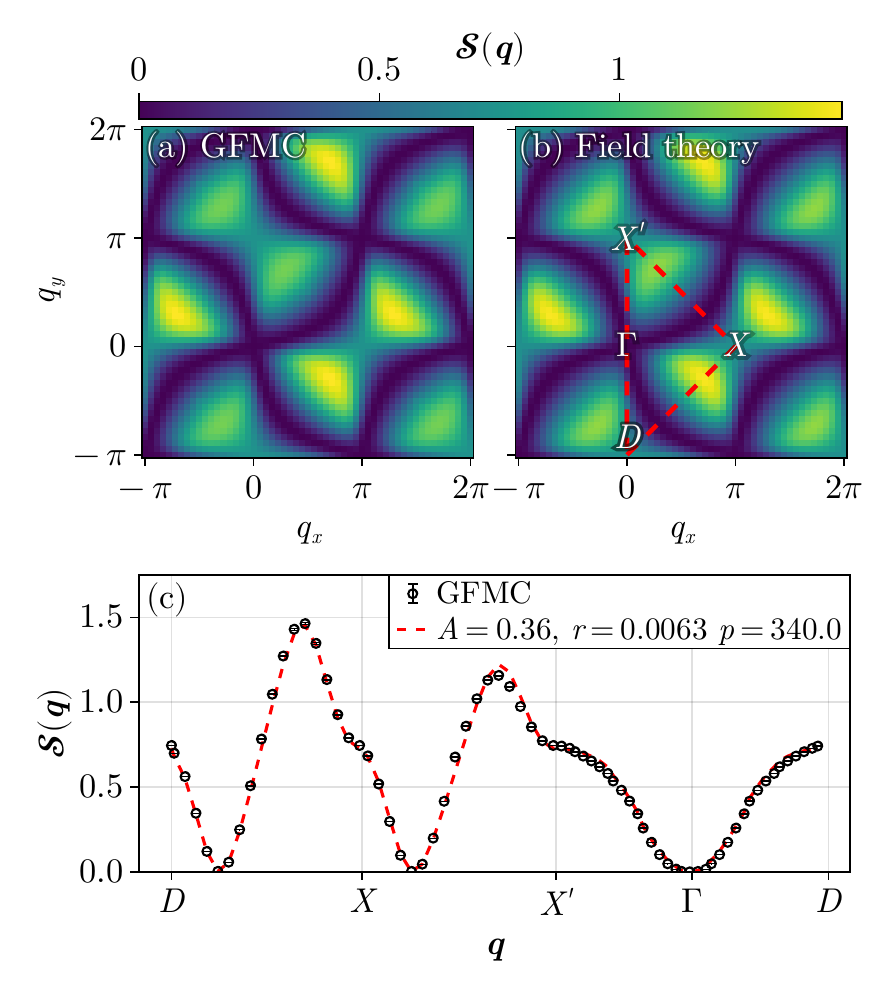}
    \caption{(a) Ground state spin structure factor in the diagonal stripe sector obtained from GFMC at $\mu=0.9J'$ for $L=36$. (b) Spin structure factor for the best fit to (a) using the asymmetric rank-2 U(1) field theory of Eqs.~(\ref{eq:maxwell}) and (\ref{eq:asymmetric}). (c) Spin structure factor along the path indicated in (b) for the spin model (black empty circles) from (a) compared to the field theory fit (red dashed line) from (b). Statistical errors are estimated from the standard deviation over 14 independent simulations.}
    \label{fig:Stair_Sq_FT}
\end{figure}

\Cref{fig:Stair_Sq_FT}(a) and (b) show the comparison of the spin structure factors from GFMC and from the best fit to the asymmetric field theory, demonstrating nearly perfect agreement. To highlight the excellent quantitative nature of this agreement, in \cref{fig:Stair_Sq_FT}(c) we show the spin structure factor for a path along high-symmetry points in reciprocal space, as indicated in \cref{fig:Stair_Sq_FT}(b). Crucially, this agreement includes the particular shape of the suppression of the pinch point given by the quadratic photon dispersion around the $\Gamma$-point, shown in \cref{fig:Condensate_mu_Sweep_xi}(d). In the Supplementary Material we show further evidence that the form of the suppression indeed precisely matches the field theory expectation $\mathcal{S}({\bm q}) \sim {\bm q}^2$. On the other hand, the agreement becomes worse when long-range order sets in. We also note that the first order confinement transition which we observe in Fig.~\ref{fig:Stair_Sq} is compatible and analogous to the first-order confinement transition known from conventional compact U(1) gauge theories in 3+1 dimensions~\cite{Jersak1983,Campos1998,Vettorazzo2004}.

\subsection{Excited sectors}
\begin{figure*}[t]
    \centering
    \includegraphics[width=\linewidth]{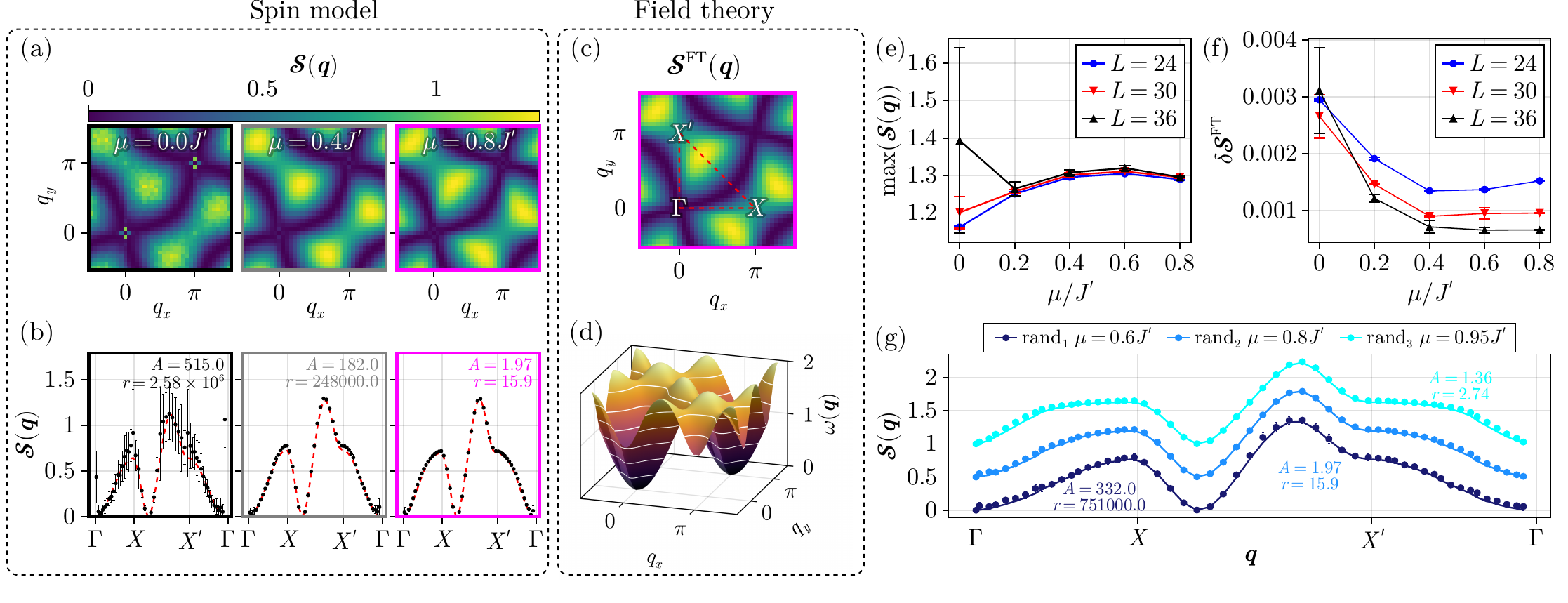}
    \caption{(a) Spin structure factor in the $6\times6$ sector [see Fig.~\ref{fig:SectorEnergies}(c)] for three different values of $\mu$ for $L=36$. (b) Data of (a) along line cuts connecting high-symmetry points in momentum space [illustrated in (c)]. The red dashed line indicates the best fit to the field theory. (c) Best fit to the field theory for the spin structure factor at $\mu=0.8J'$ shown in (a). (d) Photon dispersion as predicted by the field theory for $U=K=1$ and $W=0$. (e) Maximum of the spin structure factor as a function of $\mu$ for the $6\times6$ sector and (f) relative deviation $\delta \mathcal{S}^{\mathrm{FT}} ({\bm q})$ to the best fit to the field theory. (g) Line cuts of the spin structure factor in the randomized sectors shown in \cref{fig:SectorEnergies}(c) for different values of $\mu$. Solid lines indicate the best fit to the field theory. For visual clarity, the curve for the rand$_2$ (rand$_3$) sector is vertically shifted by $0.5$ (by $1$). In all panels, error bars are estimated by the standard deviation over 14 independent runs.}
    \label{fig:Condensate_mu_Sweep_xi}
\end{figure*}
The last section provided strong evidence that the spin-1 spiderweb model hosts an extended rank-2 U(1) QSL phase. However, its regime of stability near the RK point is relatively narrow and its energetic separation to other sectors is small, see Fig.~\ref{fig:SectorEnergies}(b). Here, we demonstrate that rank-2 U(1) QSLs in the spin-1 spiderweb model exist in much broader ranges: We also find them in generic excited sectors, where their stability region in $\mu$ is considerably larger and where they appear in $90^\circ$ rotation symmetric form. Extending our investigation to excited sectors also has a direct physical motivation. Due to the strong Hilbert space fragmentation the true ground state is hardly accessible by any conventional (numerical or experimental) annealing process such that the system would come to a rest in excited sectors. Moreover, synthetic platforms such as Rydberg atom arrays~\cite{browaeysManybodyPhysicsIndividually2020,giudiciDynamicalPreparationQuantum2022,Labuhn2016} facilitate the initialization of the system in a given classical state that may be evolved within its Hilbert space sector.

We start investigating the sector with the $6\times 6$ periodic parent state shown in Fig.~\ref{fig:SectorEnergies}(c). The reason for choosing a $6\times 6$ state is to complicate the formation of $4\times4$ ground state order, with the intent to increase the stability region of the spin liquid.
\Cref{fig:Condensate_mu_Sweep_xi}(a) shows the full momentum dependence of the spin structure factor. At $\mu=0$, where quantum fluctuations are weakest, we find the system to be highly non-ergodic, leading to relatively large statistical uncertainties, despite considerable computational effort. The structure factor $\mathcal{S}({\bm q})$ displays sharp peaks at the smallest finite resolved momenta ${\bm q} = (\pi/L,\pi/L)$. This is indicative of phase separation effects over large distances and with slow dynamics, which is the system's response to the frustration caused by the $6\times6$ parent state (further explained in the Supplementary Material). For values $\mu \geq 0.2J'$, ergodicity in this sector is restored and suppressed pinch points without any ordering peaks are observed. In contrast to the ground state sector, these patterns show intact fourfold rotational symmetry. In \cref{fig:Condensate_mu_Sweep_xi}(b) we plot $\mathcal{S}({\bm q})$ along a path of high-symmetry points in the Brillouin zone [illustrated in \cref{fig:Condensate_mu_Sweep_xi}(c)], showing again excellent agreement with the best fit to the field theory evidencing a rank-2 U(1) QSL.

To further substantiate the system's evolution into a QSL phase, in \cref{fig:Condensate_mu_Sweep_xi}(e) we show the maximum of the spin structure factor as a function of $\mu$. While the infinite unit cell in the phase-separated regime prevents conventional approaches to locate the phase boundary, we observe that for $\mu \geq 0.2J'$, the maximum becomes nearly independent of the system size indicating the absence of long-range order. Furthermore, we assess the quality of the fit to the field theory [with spin structure factor $\mathcal{S}^{\textrm{FT}}({\bm q})$] via the relative deviation $\delta \mathcal{S}^{\mathrm{FT}} = \sum_{\bm{q} \in \textrm{EBZ}}\frac{\abs{\mathcal{S}({\bm q})-\mathcal{S}^{\mathrm{FT}}({\bm q})}}{N_\textrm{sites}}$ in the extended Brillouin zone ($\mathrm{EBZ}$) plotted in \cref{fig:Condensate_mu_Sweep_xi}(f). These results also corroborate a transition into a rank-2 U(1) QSL: For $\mu \geq 0.2J'$, the fit becomes increasingly accurate ($\delta \mathcal{S}^{\mathrm{FT}}({\bm q})< 0.002$), and in particular the error decreases monotonically with system size. 

Lastly, we examine three completely generic excited sectors obtained by randomly generating fracton-free configurations [see \cref{fig:SectorEnergies}(c)] which are not associated with any periodic parent state. We refer the reader to Ref.~\cite{Niggemann2025a} for details regarding the stochastic sampling of these fracton-free configurations. The perfect agreement of the spin structure factor with the prediction of the field theory is also evident in these sectors, as shown \cref{fig:Condensate_mu_Sweep_xi}(g) illustrating line cuts of $\mathcal{S}({\bm q})$ for the three sectors at different $\mu \geq 0.6J'$.

This observation lets us conclude that the presence of an extended rank-2 U(1) QSL phase is a general feature of low-lying excited sectors. 

\section{Discussion}
The previous sections have established the spin-1 spiderweb model as a platform to realize an extended gapless fracton QSL phase in its ground state and in excited Hilbert space sectors.
In this phase, the spin structure factor obtained with error-controlled GFMC matches the predictions of a quantum rank-2 U(1) field theory with extreme precision.
In particular, we directly observe the existence of fourfold pinch points, known signatures of gapless fracton phases \cite{Yan2023_1,Yan2023_2,Benton2021,Hart2022}, and their suppression away from the Rokhsar-Kivelson point. As predicted by our field theory, we find this suppression to conform to the quadratic dispersion of the emergent photon down to small momenta where it becomes gapless. 

The numerically confirmed validity of the effective field theory in Eq.~(\ref{eq:maxwell}) indicates that the conditions under which it is constructed [see discussion below Eq.~(\ref{eq:maxwell})] are fulfilled in the spin-1 case.
Indeed, the assumption of small $B_{\sublO}$ fluctuations is known to be non-trivial in two spatial dimensions since phase-slip events $B_{\sublO}\rightarrow B_{\sublO}+2\pi$ in 2+1 dimensional spacetime, so-called \emph{instantons}~\cite{Polyakov1977}, can proliferate, gap out the photons and drive a system into an ordered phase (as is the case for the spin-1/2 spiderweb model~\cite{Niggemann2025a}). Our spin-1 model, however, shows no indications for such phenomena. Although an extremely small photon gap and weak order can never be fully excluded -- as is the case with any numerical investigation of finite-size systems -- small but finite temperatures could overcome such effects and still realize an effective rank-2 U(1) QSL. The reason why our model apparently evades the instanton effect can possibly be traced back to the absence of Lorentz-invariance. As a consequence, possible instanton configurations are highly anisotropic in 2+1 dimensional spacetime and might have the shape of world lines with actual particle-like properties (usually called visons) rather than point defects. While an in-depth discussion of such effects is beyond the scope of this work, it is important to note that the field theory is agnostic to the Hilbert space fragmentation of the spiderweb model from which it is derived. This enables the realization of a rank-2 U(1) gauge structure in a multitude of sectors, as we have demonstrated numerically.

From an experimental perspective, Rydberg atom arrays provide a particularly promising platform for realizing our model. In contrast to solid-state implementations, these systems offer a high degree of control over geometry and interactions, and square-lattice arrangements as required for the spiderweb model are routinely realized~\cite{Labuhn2016,Ebadi2021,Scholl2021}. In fact, Rydberg atom arrays have already gained attention for simulations of other lattice gauge theories, such as $\mathds{Z}_2$ spin liquids~\cite{Semeghini2021}. In our case, it is sufficient to directly engineer only the classical Ising part $\mathcal{H}_1$: once the associated higher-rank Gauss law constraint is implemented, weak quantum fluctuations (e.g., via laser-induced transverse fields) will generically generate the dynamical terms contained in $\mathcal{H}_2$ perturbatively.

The main challenge lies in engineering the required pattern of interaction strengths beyond the natural van-der-Waals form $\sim 1/r^6$. However, several recent proposals demonstrate that substantial tunability of effective interactions can be achieved. In particular, Floquet schemes based on periodic modulation of local Rydberg states~\cite{tian2025} and multicolor Rydberg dressing protocols~\cite{Wu2022} allow for programmable interaction profiles that can, in principle, approximate the coupling structure required in $\mathcal{H}_1$. Another important ingredient is the realization of effective spin-1 degrees of freedom. This can be achieved using multi-level encoding in Rydberg systems using multiple Rydberg states~\cite{Mogerle2025,robertQuditEncodingRydberg2025,liu2024}. We also note that our finding of fracton spin liquids in excited sectors will aid experiments, which may be prohibited from reaching the true ground state due to Hilbert space fragmentation and glassy dynamics, which are generic features of fracton models.

\section{Methods}

    \subsection{Green function Monte Carlo}\label{app:gfmc}
    Quantum Monte Carlo methods are numerically exact ways to determine ground state properties of so-called \emph{stochastic Hamiltonians}. Here, we discuss our implementation of the Green function Monte Carlo (GFMC) approach as detailed in Ref.~\cite{Buonaura1998}, which we find particularly useful to our given problem, see also Refs.~\cite{becca2017quantum,Trivedi1990,Hetherington1984}. In essence, GFMC avoids exhaustive computations in the entire Hilbert space in favor of a random walk, i.e. a Markov-chain, to facilitate statistical sampling of observables.
    The random walk is utilized to realize a projection approach, i.e. projecting out excited states from a trial state $\ket{\psi_\textrm{T}}$ with finite overlap to the ground state $\ket{\psi}$. One such projection is given in terms of the Hamiltonian $\mathcal{H}$ as $(\Lambda - \mathcal{H})^\p \ket{\psi_\textrm{T}} \rightarrow \ket{\psi}$ , where $\Lambda>0$ is a constant and $\p \rightarrow \infty$ the projection order. If $\mathcal G \equiv \Lambda - \mathcal{H}$ has only non-negative elements it can be expressed through a Markovian matrix, whose elements define normalized probabilities of transitioning from one classical state to another, allowing the usage of a Monte Carlo approach. While the constant $\Lambda$ can enforce positivity of the diagonal elements, the so-called \emph{sign problem} arises if the Hamiltonian has positive off-diagonal elements.
    In the present work, we employ the continuous time-limit modification of the method presented in Ref.~\cite{ralkoDynamicsQuantumDimer2006}. This approach performs the exact limit $\Lambda \rightarrow \infty$ in which case the projector is exactly equal to an imaginary time evolution operator $\e^{-\mathcal{H}\Delta \tau}$, where $\Delta \tau$ is a chosen time-step. Note that on a lattice, no trotterization error occurs regardless of the size of $\Delta \tau$.
    The requirement $\braket{\psi}{\psi_\textrm{T}} = \sum_x \braket{\psi}{x}\braket{x}{\psi_\textrm{T}} \neq 0$ can be seen to be satisfied by choosing $\psi_\textrm{T}(x)\equiv \braket{x}{\psi_\textrm{T}} > 0$ for all configurations $\ket{x} = \ket{S^z_1,S^z_2,\dots,S^z_{N_\textrm{sites}}}$.
    
    \emph{Importance sampling --}
    Importance sampling is implemented through the trial function itself, which is therefore also referred to as the \emph{guiding wavefunction}. Starting from a spin configuration $\ket{x} = \ket{S^z_1,S^z_2,\dots,S^z_{N_\textrm{sites}}}$, each Markov step consists of sampling a new configuration with a probability proportional to the weight $\mel{x}{\mathcal{G}}{x'} \frac{\psi_\textrm{T}(x')}{\psi_\textrm{T}(x)}$.
    This new configuration then contributes to averages of observables via an accumulation $\prod_{i=n}^{n+\mathcal{P}} w_{x_i}$ of the total weight $w_x = \sum_{x'}\mel{x}{\mathcal{G}}{x'} \frac{\psi_\textrm{T}(x')}{\psi_\textrm{T}(x)}$ over the previous steps.
    This summation can be evaluated efficiently for local Hamiltonians where the number of elements $\mathcal{H}_{xx'} = \mel{x}{\mathcal{H}}{x'}$ scales linearly in system size for a given $\ket{x}$. 
    The configurations drawn from the Markov chain this way will be distributed according to an equilibrium condition that is determined by $\mathcal{G}$ and $\psi_\textrm{T}(x)$.

    Importantly, if $\ket{\psi_\textrm{T}}$ is equal to the exact ground state, the variance of the energy is zero and the projection converges at zeroth order. On the other hand, observables which do not commute with the Hamiltonian have a finite variance even if the exact ground state is used, although this variance will generally be larger for less precise wave functions. 
    
    While the exact wavefunction is only known at the RK-point, fluctuations and rate of projection convergence are still greatly improved if the guiding wavefunction is reasonably close to the ground state. Here, we choose a standard Jastrow ansatz for the guiding wave function, which captures two-body correlations up to arbitrary length
    \begin{equation}
        \log \psi_\textrm{T}(x) = \sum_i m_i x_i + \frac{1}{2}\sum_{ij} V_{ij} x_i x_j,
    \end{equation}
    where $m_i$ and $V_{ij} =V_{ji}$ are real variational parameters that are optimized using the stochastic reconfiguration method \cite{becca2017quantum} before the start of the GFMC simulation.

    Clearly, the Jastrow ansatz captures the exact RK- wavefunction $\psi_\textrm{RK}(x) = 1$, and is efficient to evaluate numerically.
    During the Markov chain we only need to evaluate the ratio of wavefunction amplitudes between two configurations that differ by a single cluster flip, i.e. $\ket{x'} = \mathcal{F}_{\sublO} \ket{x}$, yielding
    \begin{equation}
        \log \left( \frac{\psi_\textrm{T}(x')}{\psi_\textrm{T}(x)}\right) = \sum_i m_i(x'_i-x_i) + \frac{1}{2}\sum_{ij} V_{ij}(x'_ix'_j-x_ix_j).
    \end{equation}
    A single cluster flip only affects eight sites which we label by $\sublO_k \equiv \sublO_1,\dots,\sublO_8$. Using $V_{ij} = V_{ji}$, denoting the change in the spin configuration as $F_k \in \pm 1$ and further defining the effective local field $h_{i} = m_i + \sum_j V_{ji} x_j$, we may express this ratio as
    \begin{equation}
        \log \left( \frac{\psi_\textrm{T}(x')}{\psi_\textrm{T}(x)}\right) = \sum_{k=1}^8  h_{\sublO_k} F_k + \frac{1}{2}\sum_{kk'} V_{\sublO_k \sublO_{k'}}F_kF_{k'}
    \end{equation}  
    which may be computed in $\mathcal{O}(1)$ complexity, independent of the system size \footnote{While $h$ contains a sum over all sites, it only needs to be updated after each move as $h_j \rightarrow h_j + \sum_k V_{i_k,j} F_k$, while the ratio of wavefunctions needs to be evaluated $\sim L\times L$ times before each move.}. 

    Due to this efficient evaluation, we found the Jastrow function to outperform more expressive wavefunction approaches (which can describe a larger manifold of wavefunctions) such as restricted Boltzmann machines~\cite{Carleo2017,He-Yu2024}.

    \emph{Many walker formalism --}
    To reduce statistical fluctuations, particularly at large projection times $\tau$, we employ the many walker formalism as introduced in Ref. \cite{Buonaura1998}. This approach propagates a population of walkers independently of each other for a few steps $n_\textrm{branch}$, (or, a small imaginary time $\Delta \tau \sim 0.1$), during which they accumulate their weight as $\prod_{n=1}^{n_\textrm{branch}}w_{x_n}$ after which they may recombine, ensuring each walker's survival with probability proportional to their accumulated weight. A larger number of walkers will thus explore the Hilbert space more efficiently, leading to a lower variance for larger projection times as well as a more rapid convergence of the projection scheme. 

    We note that unlike other walker population control mechanisms  \cite{Runge1992,Hetherington1984} or in \emph{diffusion Monte Carlo}, no systematic bias is introduced in this approach regardless of the number of walkers~\cite{becca2017quantum}. 
    
    \emph{Ergodicity--}
    While error controlled, GFMC as any Markov chain method may suffer from poor ergodicity. A good diagnostic tool is to compute errors using the standard deviation of several fully independent runs, i.e. initializing the walkers with randomized configurations. For problems with poor ergodicity, the error obtained this way can be significant, as visible in the leftmost panel of \cref{fig:Condensate_mu_Sweep_xi}(b).
    In the present case uniformly sampling from a single sector of the fragmented Hilbert space is not possible.
    Instead, we first initialize $N_w \sim 20{,}000$ walkers in a given sector by specifying the initial configuration and subsequently perform a long series of $\gtrsim 10^7$ fully random cluster flips. This leads to a state of the walker ensemble which is uncorrelated with the initial configuration.

    We emphasize that this procedure does not improve the ergodicity of the Markov chain, but rather serves as a diagnostic tool to provide accurate error estimates.

\section*{Additional information}
\noindent
\textbf{CORRESPONDENCE} and requests for materials should be addressed to
Nils Niggemann.

\noindent
\textbf{CODE AVAILABILITY}
Code used to perform the numerical simulations presented in this work is openly accessible in the GitHub repository \url{https://github.com/NilsNiggemann/SpiderWebModel.jl}.

\noindent
\textbf{DATA AVAILABILITY}
Simulation data are available at \url{https://zenodo.org/records/16781381}.

\textbf{ACKNOWLEDGMENTS}
We are grateful to Björn Sbierski for providing helpful comments on this manuscript. Further, we are grateful towards Subir Sachdev, Arnaud Ralko, Karlo Penc, Arnab Sen, Robin Schäfer, Daniel Lozano-Gómez, Han Yan, Federico Becca, Yuan Wan, Alaric Sanders, Rhine Samajdar and Marcello Dalmonte for helpful discussions.

\textbf{FUNDING STATEMENT}
N.~N.~and J.~R.~acknowledge support from the Deutsche
Forschungsgemeinschaft (DFG, German Research Foundation), within Project-ID 277101999 CRC 183 (Project
A04).
N.~N.~ further acknowledges funding from the European Research
Council (ERC) under the European Union’s Horizon ERC-2022-COG Grant with Project number 101087692.
M.~A.~ acknowledges funding from the PNRR MUR project PE0000023-NQSTI.
We acknowledge the use of the JUWELS cluster at the Forschungszentrum J\"ulich and the Noctua2 cluster at the Paderborn Center for Parallel Computing (PC$^2$).

\noindent
\textbf{AUTHOR CONTRIBUTIONS}
The project was under the supervision of J.~R.~ J.~R.~ and N.~N.~ conceived the project. N.~N.~ and M.~A.~ carried out calculations regarding the classical properties of the model and N.~N.~ planned and performed the numerical investigation of the quantum model. J.~R.~ and Y.~S.~ carried out calculations for the solution of the gauge theory. N.~N.~ wrote the first draft of the paper. All authors contributed to the revision of the paper.

\noindent
\textbf{COMPETING INTERESTS}
The authors declare no competing interests.

\bibliography{bib}

\begin{appendices}
\newpage
\section{Derivation of the ground state constraint}\label{app:constraint_derivation}

Here, we briefly summarize the derivation of the ground state constraint $\mathcal{C}_{\sublX}=0$, found in more detail in Ref.~\cite{Niggemann2025a}.
We start with the continuum form of the chargeless rank-2 Gauss law, where $E^{\mu\nu}$ is the traceless, symmetric rank-2 electric field and $\mu,\nu\in\{x,y\}$ such that
\begin{align}
  \label{eq:Gauss_law}
  (\partial^2_x - \partial^2_y) E^{xx}({\bm r}) + 4\partial_x\partial_y E^{xy}({\bm r}) = 0
\end{align}
In principle, in a rotation-invariant form of the Gauss law $\partial_\mu\partial_\nu E^{\mu\nu}=0$ a factor 2 instead of a factor 4 would appear in the second term of Eq.~(\ref{eq:Gauss_law}). However, this would lead to non-uniform prefactors in the corresponding spin constraint and we therefore prefer to use the constraint in Eq.~(\ref{eq:Gauss_law}). We begin by discretizing the derivatives over unit vectors $\hat{x},\hat{y}$, for example $\partial^2_x E^{xx}({\bm r}_i) \rightarrow E^{xx}({\bm r}_i+\hat{x}) - 2E^{xx}({\bm r}_i) +E^{xx}({\bm r}_i-\hat{x})$.
In discretized form, \cref{eq:Gauss_law} then becomes
\begin{align}
0 = +&E^{xx}({\bm r}_i+\hat{x}) + E^{xy}({\bm r}_i+\hat{x}+\hat{y})\nonumber\\
- &E^{xx}({\bm r}_i+\hat{y}) - E^{xy}({\bm r}_i+\hat{y}-\hat{x}) \nonumber\\
  +&E^{xx}({\bm r}_i-\hat{x}) + E^{xy}({\bm r}_i-\hat{y}-\hat{x})\nonumber\\
  -&E^{xx}({\bm r}_i-\hat{y})- E^{xy}({\bm r}_i+\hat{x}-\hat{y}) \label{eq:constraint} 
\end{align}
In order to map this equation to a spin model, we associate the two inequivalent components with independent and unconstrained spin degrees of freedom, each localized on a different sublattice of a square lattice in a unit cell at position ${\bm r}_i$.
\begin{align}
  E^{xx}({\bm r}_i) &= -E^{yy}({\bm r}_i) = S^z_{\sublO}({\bm r}_i) \label{eq:Exx_Sz}\\
  E^{xy}({\bm r}_i) &= E^{yx}({\bm r}_i) = S^z_{\sublX}({\bm r}_i)\label{eq:Exy_Sz}
\end{align}
By inserting \cref{eq:Exx_Sz,eq:Exy_Sz} into \cref{eq:constraint}, we arrive at the constraint $\mathcal{C}_{\sublX}=0$ introduced in the main text.

This construction guarantees the existence of an emergent classical rank-2 U(1) gauge theory in the Gaussian approximation of unconstrained spin degrees of freedom.

\section{Number of classical ground states}\label{app:ground_states}
We derive an estimate for the dimension $n$ of the constrained Hilbert space in which $\mathcal{C}_{\sublX}=0$ for all eight-site clusters centered around $\sublX$. In particular, for a system with $N_{\textrm{sites}}$ spins we express $n$ as $n=b^{N_\textrm{sites}}$ and calculate the base $b$.

A simple estimate, following Pauling counting arguments, approximates $n$ as $n\approx3^{N_\textrm{sites}} k^{N_\textrm{sites}/2}$ with the total number of spin-1 Ising configurations given by $3^{N_\textrm{sites}}$ and defining $k$ as the ratio between the number of all configurations satisfying $\mathcal{C}_{\sublX}=0$ for one cluster $\sublX$ and the total number of $3^8$ Ising configurations in one cluster. Furthermore, $N_\textrm{sites}/2$ is the number of constraints. This approximation gives $b\approx\sqrt{41/27}=1.232$.

More rigorously, we may also determine $b$ by systematically calculating upper and lower bounds. For estimating an upper bound of $b$, we consider systems with periodic boundaries and $N_\textrm{sites}=L\times L$ sites. Specifically, we determine $n$ for $L=4$, $L=6$, and $L=8$ by exhaustively generating all states in the constrained subspace. We note that for $L>8$ this becomes numerically too difficult due to the large number of states. The obtained values are given in Table~\ref{tab:numGS} together with the base $b$ calculated through $b=n^{1/L^2}$. The base $b$ decreases monotonically in $L$ for $L=4,6,8$, and therefore the value $b=1.468$ for $L=8$ can be used as an upper bound.

A lower bound of $b$ is found by determining the exact number of states that are obtained by applying fluctuator moves in the configuration of Fig.~\ref{fig:construction_sectors}(a). This state has a $\sqrt{10}\times\sqrt{10}$ unit cell indicated by red dashed lines and it has one flippable cluster per unit cell (their centers are marked by red dots). Importantly, these flippable clusters do not overlap such that they can be flipped independently. Furthermore, under the condition that flippable clusters do not overlap, their density (one flippable cluster per 10 sites) is maximal in this state. Since each flippable cluster can be in three different configurations, fluctuator moves starting from Fig.~\ref{fig:construction_sectors}(a) can generate $3^{N_\textrm{sites}/10}$ different states in the constrained subspace, yielding the lower bound $b=3^{1/10}=1.116$. In total, one finds
\begin{align}
b=1.292&\pm0.176, \label{eq:hilbert_space_size}
\end{align}
where the error margins range upward (downward) to the upper (lower) bounds, which agrees with the Pauling estimate. We note that in the case of spin-$1/2$, the Pauling estimate is found to be surprisingly inaccurate; see Ref.~\cite{Niggemann2025a} for details.

\begin{table}
\begin{tabular}{|c|c|c|}
\hline
& Number of states $n$ & $b=n^{1/L^2}$\\\hline
$L=4$ & 6 859 & 1.737\\
$L=6$ & 5 800 827 & 1.541\\
$L=8$ & 47 067 992 003 & 1.468\\
\hline
\end{tabular}
\caption{Dimension $n$ of the constrained subspace for a system with periodic boundaries and $N_\textrm{sites}=L\times L$ sites, where $L=4$, $L=6$, and $L=8$. The third column shows the base $b$ defined by $n=b^{N_\textrm{sites}}$.}\label{tab:numGS}
\end{table}

\section{Hilbert space fragmentation}\label{app:fragmentation}
The dynamics generated by the fluctuators $\mathcal{F}_{\sublO}$ and $\mathcal{F}_{\sublO}^\dagger$ does not cover the full constrained subspace of the spin-1 spiderweb model, but splits it into many dynamically disconnected sectors. This property, referred to as Hilbert space fragmentation, is known from other fracton models~\cite{Sala2020,khudorozhkovHilbertSpaceFragmentation2022,stahlStrongHilbertSpace2025,Adler2024,Will2024,Feng2022,leeFrustrationinducedEmergentHilbert2021}. Here, we prove the existence of Hilbert space fragmentation for the spiderweb model, i.e., that the number of dynamically disconnected sectors grows exponentially with the number of sites. While we focus on the case of spin-1, we note that this argument holds for any spin, in particular also spin-$1/2$; see Ref.~\cite{Niggemann2025a}. 

\begin{figure}
    \centering
    \includegraphics[width = 0.9\linewidth]{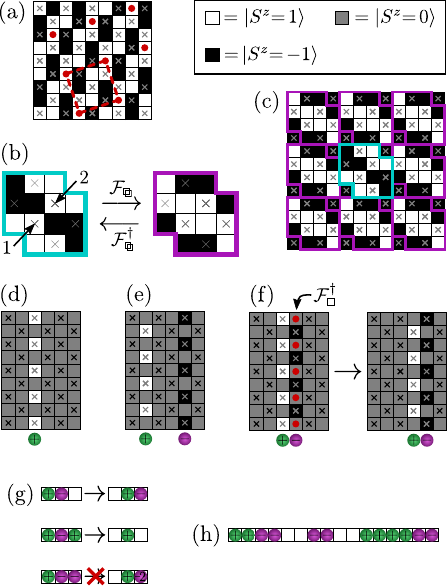}
    \caption{(a) Spin configuration with the largest density of non-overlapping flippable eight-site clusters, to estimate a lower bound for the dimension of the constrained subspace. (b) Definition of the double fluctuator $\mathcal{F}_{\includegraphics[scale=0.2]{double_move.pdf}}$. Note that the two sites 1 and 2 are not flipped by $\mathcal{F}_{\includegraphics[scale=0.2]{double_move.pdf}}$. (c) $4\times4$ periodic state where the action of $\mathcal{F}_{\includegraphics[scale=0.2]{double_move.pdf}}^\dagger$ in any unit cell leads to a new Hilbert space sector. (d) Single $S^z$-string charge. (e) Single $S^z$-string dipole. (f) Elementary string-dipole move generated by applying the fluctuator $\mathcal{F}_{\sublO}^\dagger$ on the eight-site clusters around the red dots. (g) Rules for string-dipole moves. While the top and middle moves are allowed, the process at the bottom is forbidden due to the spin-1 constraint $|S_i^z|\leq1$. (h) Example of a sequence of paired string charges for proving the existence of subextensively many Hilbert space sectors from string states in the spin-1 system. }\label{fig:construction_sectors}
\end{figure}

For our proof we consider a new type of fluctuator $\mathcal{F}_{\includegraphics[scale=0.2]{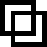}}$ shown in Fig.~\ref{fig:construction_sectors}(b). Its action on a 14-site cluster flips 12 of these spins by two units of angular momentum (i.e. locally acts as $S_i^+ S_i^+$ or $S_i^- S_i^-$) but does not change the two center spins [marked with numbers 1 and 2 in Fig.~\ref{fig:construction_sectors}(b)]. Importantly, $\mathcal{F}_{\includegraphics[scale=0.2]{double_move.pdf}}$ is defined in a way that it annihilates states where the center spins 1 and 2 are not equal or when they are both in $S^z=0$ configurations.  This may be enforced, via the simple projector $ \left[1 - \frac{1}{4}\left(S_1^z- S_2^z\right)^2\right] \times \left(S_1^z \right)^2 \left(S_2^z \right)^2$. In the illustrated case, the center spins have $S^z=1$, however, our arguments also apply to the case where they are in $S^z=-1$ configurations.

If one only considers the net changes $\Delta S_i^z$ at these 14 sites, the fluctuator $\mathcal{F}_{\includegraphics[scale=0.2]{double_move.pdf}}$ is identical to $\mathcal{F}_{\sublO}\mathcal{F}_{\sublO}\mathcal{F}_{\sublO'}\mathcal{F}_{\sublO'}$ acting on two eight-site clusters centered around the sites $\sublO$ and $\sublO'$ ($\sublO$ and $\sublO'$ are those sites that both have the sites 1 and 2 as their nearest neighbors). Therefore, $\mathcal{F}_{\includegraphics[scale=0.2]{double_move.pdf}}$ and $\mathcal{C}_{\sublX}$ commute for all $\sublX$ which means that $\mathcal{F}_{\includegraphics[scale=0.2]{double_move.pdf}}$ operates {\it within} the constrained subspace. However, if one now considers the consecutive execution of the four operators $\mathcal{F}_{\sublO}$, $\mathcal{F}_{\sublO}$, $\mathcal{F}_{\sublO'}$, $\mathcal{F}_{\sublO'}$ one finds that there is no order in which these operators can be applied such that they act like $\mathcal{F}_{\includegraphics[scale=0.2]{double_move.pdf}}$, since they always annihilate the state in Fig.~\ref{fig:construction_sectors}(b). This is due to the condition that the two center spins 1 and 2 fulfill $S_1^z\neq0$, $S_2^z\neq0$ and $S_1^z=S_2^z$. In other words, the action of the double fluctuator $\mathcal{F}_{\includegraphics[scale=0.2]{double_move.pdf}}$ on these 14 sites cannot be expressed as the sequential action of $\mathcal{F}_{\sublO}$ and $\mathcal{F}_{\sublO'}$. Consequently, the two 14-site states in Fig.~\ref{fig:construction_sectors}(b) are in different Hilbert space sectors under the action of $\mathcal{F}_{\sublO}$ (and $\mathcal{F}^\dagger_{\sublO}$).

The property that $\mathcal{F}_{\includegraphics[scale=0.2]{double_move.pdf}}$ creates new Hilbert space sectors often no longer holds if the motif in Fig.~\ref{fig:construction_sectors}(b) is part of a larger system with more than 14 sites. This is because in a larger system it may be possible to express $\mathcal{F}_{\includegraphics[scale=0.2]{double_move.pdf}}$ as
\begin{equation}
\mathcal{F}_{\includegraphics[scale=0.2]{double_move.pdf}}=\cdots\mathcal{F}_{\sublO}\cdots\mathcal{F}_{\sublO}\cdots\mathcal{F}_{\sublO'}\cdots\mathcal{F}_{\sublO'}\cdots,\label{eq:sequential_F}
\end{equation}
where ``$\cdots$'' denotes fluctuator moves $\mathcal{F}$ on other surrounding eight-site clusters. To prove the exponential scaling of the number of Hilbert space sectors, it is, however, sufficient to find {\it one} periodic state with the motif in Fig.~\ref{fig:construction_sectors}(b), in which $\mathcal{F}_{\includegraphics[scale=0.2]{double_move.pdf}}$ cannot be expressed via elementary fluctuators $\mathcal{F}_{\sublO}$ as in Eq.~(\ref{eq:sequential_F}). Such a configuration is shown in Fig.~\ref{fig:construction_sectors}(c), which has a $4\times4$ unit cell containing the motif in Fig.~\ref{fig:construction_sectors}(b). Another configuration with this property is discussed in Ref.~\cite{Niggemann2025a}. Note that the action of the double fluctuator $\mathcal{F}^{\dagger}_{\includegraphics[scale=0.2]{double_move.pdf}}$ is illustrated in the center $4\times4$ unit cell of Fig.~\ref{fig:construction_sectors}(c). Importantly, none of the states obtained by applying $\mathcal{F}^{(\dagger)}_{\includegraphics[scale=0.2]{double_move.pdf}}$ in different unit cells have eight-site clusters that can be flipped by $\mathcal{F}_{\sublO}$ or $\mathcal{F}^\dagger_{\sublO}$. Consequently, the action of $\mathcal{F}_{\includegraphics[scale=0.2]{double_move.pdf}}^\dagger$ in any $4\times4$ unit cell leads to a new Hilbert space sector. A lower bound for the number of sectors can now be estimated as $2^{N_\textrm{sites}/16}$, demonstrating its exponential scaling with system size. However, note that each of these sectors is trivial in the sense that it contains only a single state.

We emphasize that in a generic spin-1 configuration with the 14-site motif in Fig.~\ref{fig:construction_sectors}(b), the action of $\mathcal{F}_{\includegraphics[scale=0.2]{double_move.pdf}}$ will usually {\it not} result in a new sector because $\mathcal{F}_{\includegraphics[scale=0.2]{double_move.pdf}}$ can be expressed as in Eq.~(\ref{eq:sequential_F}). In rare exceptions [Fig.~\ref{fig:construction_sectors}(c)], the creation of new Hilbert space sectors via $\mathcal{F}_{\includegraphics[scale=0.2]{double_move.pdf}}$ is only possible because the 14-site motif is trapped in an environment with inactive or severely limited $\mathcal{F}_{\sublO}$-dynamics (e.g. due to the absence of local $S_i^z=0$ states). Stated differently, we could not find a spin-1 state where the application of $\mathcal{F}_{\includegraphics[scale=0.2]{double_move.pdf}}$ creates a new sector while the $\mathcal{F}_{\sublO}$-dynamics in these sectors is connected and can spread over the whole system. We could generally not identify any mechanism to construct exponentially many Hilbert space sectors that have non-trivial dynamics. We therefore conclude that while spin states like in Fig.~\ref{fig:construction_sectors}(c) can be used to prove the existence of exponentially many sectors, they are highly artificial and the total number of states they contain covers only a small part of the constrained Hilbert space.

On the other hand, there exists a way to construct subextensively many sectors (i.e., whose number scales exponentially with the linear system size $L$) that have non-trivial and collective dynamics, possibly hosting a fracton spin liquid. Their construction is based on the string-like configuration on sublattice 1 shown in Fig.~\ref{fig:construction_sectors}(d). In the following, we define the operator $\mathcal{G}_{\bm |}=\prod S^+_{\sublX}$, which creates such a string when acting on a homogeneous $S_i^z=0$ state. We assign a `string-charge' $\mathcal{Q}=\pm1$ to these states, indicating whether the strings are built from $S_i^z=1$ or from $S_i^z=-1$ states. We call such configurations `$S^z$-strings' and mark them with green balls (for $\mathcal{Q}=+1$) or violet balls (for $\mathcal{Q}=-1$) in Fig.~\ref{fig:construction_sectors}(d)-(h). The term `charge' in this context should not be confused with the fractonic matter charges $\rho$. All configurations with one or more strings are translation-invariant in the $y$ direction and fulfill all constraints $\mathcal{C}_{\sublX}=0$. The following discussion considers only such translation-invariant states and treats them as representative states for the entire Hilbert space sectors they belong to, in which $y$-translation invariance of individual states is typically broken. In a similar way, our arguments also hold for horizontal or diagonal strings (note, however, that a diagonal string has to be defined on sublattice 2 in order to respect all ground-state constraints).

The operator $\mathcal{G}_{\bm |}$ does not commute with the conserved subdimensional magnetizations $M_{\bm \diagdown}$, $M_{\bm \diagup}$ in Fig.~2 of the main text and thus cannot be constructed by repeated applications of $\mathcal{F}_{\sublO}$. Therefore, adding a $S^z$-string generates a new Hilbert space sector.
Next, we consider a string-dipole consisting of a $\mathcal{Q}=+1$ string and a $\mathcal{Q}=-1$ string, as shown in Fig.~\ref{fig:construction_sectors}(e), which is created by the operator $\mathcal{G}_{{\bm |}_1}\mathcal{G}^\dagger_{{\bm |}_2}$. Since such operators also cannot be built from individual fluctuators $\mathcal{F}_{\sublO}$, because this would require {\it fractional} fluctuator moves (see discussion in Fig.~2 of Ref.~\cite{Niggemann2025a}), adding a string-dipole again gives rise to a new Hilbert space sector.
The simplest change of one string configuration into another, which does {\it not} create a new Hilbert space sector, is the translation of a string-dipole in the perpendicular direction as shown in Fig.~\ref{fig:construction_sectors}(f). This string-dipole move can be generated by applying fluctuators $\mathcal{F}_{\sublO}^\dagger$ on the eight-site clusters around the red dots in Fig.~\ref{fig:construction_sectors}(f) and corresponds to the insertion of a string-quadrupole.

Importantly, however, this motion of string dipoles is restricted by the spin constraint $S_i^z\in\{-1,0,1\}$.
To see this, we summarize the rules for string moves in Fig.~\ref{fig:construction_sectors}(g), where we only show the effective one-dimensional system of green/violet string charges along the $x$ direction and omit the $y$ direction. The move illustrated in the top panel of Fig.~\ref{fig:construction_sectors}(g) is just the elementary dipole translation of Fig.~\ref{fig:construction_sectors}(f). The middle panel of Fig.~\ref{fig:construction_sectors}(g) also shows an allowed elementary dipole move where the `$+-$' dipole in the sequence `$+-+$' is moved to the right by one lattice spacing, involving the annihilation of a positive and negative string-charge. On the other hand, as shown in the bottom panel of Fig.~\ref{fig:construction_sectors}(g), a `$+-$' dipole cannot be moved to the right if it is blocked by another `$-$' string charge, as this would generate sites with $S_i^z=-2$. Therefore, a string dipole cannot be moved across a single string-charge by fluctuator moves $\mathcal{F}_{\sublO}$.

This blocking effect in the motion of string-dipoles due to the spin length constraint gives rise to a subextensive scaling of the number of Hilbert space sectors. To understand this and to estimate a lower bound for the number of sectors, we consider a {\it subset} of all possible string configurations in which all arrangements of string charges ($\mathcal{Q}=-1$, $\mathcal{Q}=0$ or $\mathcal{Q}=+1$) occur in nearest-neighbor pairs of equal charges [see Fig.~\ref{fig:construction_sectors}(h) for an example]. Importantly, dipole motion is completely blocked in all of these states due to the forbidden process in the bottom panel of Fig.~\ref{fig:construction_sectors}(g), such that different states cannot be transformed into each other by $\mathcal{F}_{\sublO}$ moves. Therefore, each of these paired string configurations can be regarded as a representative state in distinct Hilbert space sectors. The number of these configurations scales subextensively as $3^{L_x/2}$ where the base 3 stands for the three possible charges $\mathcal{Q}=-1$, $\mathcal{Q}=0$ or $\mathcal{Q}=+1$ and the factor $1/2$ in the exponent is due to the pairing of strings.

Crucially, in contrast to the extensively many sectors constructed via the double moves $\mathcal{F}_{\includegraphics[scale=0.2]{double_move.pdf}}$, the subextensively many sectors from string configurations usually allow for non-trivial, collective dynamics with overlapping fluctuator moves $\mathcal{F}_{\sublO}$. We expect that all states in the subextensively many string sectors altogether cover a much larger part of the constrained Hilbert space than the sectors from $\mathcal{F}_{\includegraphics[scale=0.2]{double_move.pdf}}$-moves.

\begin{figure}
  \centering
  \includegraphics[width=0.9\linewidth]{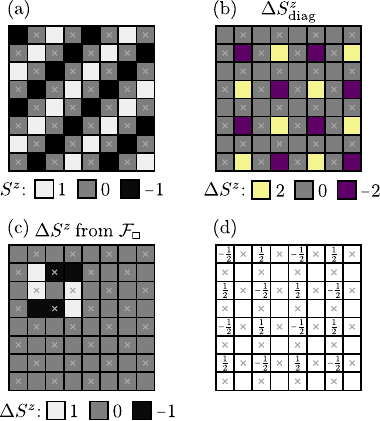}
  \caption{(a) $90^\circ$ rotated version of the diagonal stripe state in Fig.~3(a) of the main text. (b) Difference $\Delta S^z_\text{diag}$ between the diagonal stripe states in Fig.~3(a) of the main text and in (a). (c) Spin changes $\Delta S^z$ from the application of $\mathcal{F}_{\sublO}$ in one eight-site cluster. (d) Multiplicities of the applications of $\mathcal{F}_{\sublO}$ to generate $\Delta S^z_\text{diag}$ in (b).}
  \label{fig:staircase_relation}
\end{figure}

\section{Relation between fragmentation and spin quantum number} \label{app:fragmentation_spin}
\begin{figure}
    \centering
    \includegraphics[width = \linewidth]{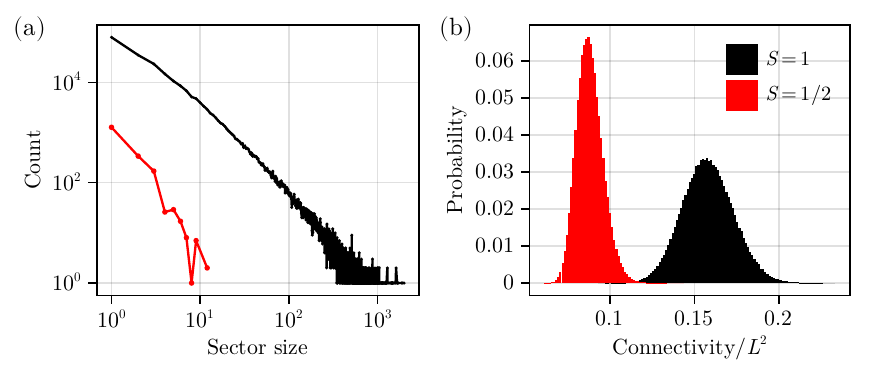}
    \caption{
    (a) Histogram of connected-sector sizes for spin-$1/2$ and spin-1 for $L=6$ obtained via exact enumeration of solutions. (b) Probability distribution of connectivity of configurations in the staircase (spin-$1/2$) and diagonal stripe (spin-1) sector for $L=28$, obtained via GFMC sampling of configurations at the RK point.}
    \label{fig:sector_hist}
\end{figure}

In Ref.~\cite{Niggemann2025a}, it was found that fragmentation has prohibitive effects on the dynamics of the spin-$1/2$ spiderweb model. 
In the main text, the effects of fragmentation were found to be qualitatively weaker for spin-1. To quantify this difference, we compare the number and sizes of sectors for spin-$1/2$ and spin-1 for different system sizes $L\times L$ with periodic boundaries by exact counting of all possible configurations. The results are summarized in \cref{tab:sectors}. 
As both the number of fracton-free configurations and sectors grow exponentially in $L^2$, it is useful to define the fragmentation coefficient 
\begin{equation}
  f= (n_\textrm{Sectors}/n_\textrm{Total})^{1/L^2}, \label{eq:fragmentation_coefficient}
\end{equation}
where $n_\textrm{Total}$ is the total number of fracton-free spin states and $n_\textrm{Sectors}$ is the number of Hilbert space sectors for a given system size $L$. If there is no fragmentation, the number of sectors grows subexponentially as $L \rightarrow \infty$, whereas $n_\textrm{Total} = b^{L^2}$ and thus $f=0$. On the other hand, for maximal fragmentation each configuration forms its own sector, $n_\textrm{Sectors}=n_\textrm{Total}$ and thus $f=1$. 
In the case of the spiderweb model, we find $f\sim0.933$ for spin-1 and $f=0.982$ for spin-$1/2$. While both are indicative of very strong fragmentation, the value for spin-1 is significantly smaller, consistent with the observation of weaker fragmentation.
While $f$ is a useful measure for the degree of fragmentation, it does not capture the effect that fragmentation has on the dynamics within individual sectors. 
To quantify this, we compare the sizes of sectors in \cref{fig:sector_hist}(a), and find that for spin-$1$, the size of Hilbert space sectors increases considerably compared to spin-$1/2$ at an equal system size of $L=6$. Another useful measure is the connectivity which specifies how many states can be reached from a starting configuration by applying a single eight-site fluctuator move. In \cref{fig:sector_hist}(b) we display the considerably larger average connectivity of configurations for spin-$1$ in the diagonal stripe sector as compared to the lowest energy sector reported in Ref.~\cite{Niggemann2025a} for spin-$1/2$ (dubbed staircase sector). Both results, respectively, were obtained by GFMC sampling of configurations at the RK point for $L=28$.
 
\begin{table}[h]
\begin{tabular}{l|cc||cc||cc}
  \toprule
  $L$ & \multicolumn{2}{c}{Total} & \multicolumn{2}{c}{Sectors} & \multicolumn{2}{c}{f} \\
  \cmidrule{2-3} \cmidrule{4-5} \cmidrule{6-7}
  $S$& \nicefrac{1}{2} & 1 & \nicefrac{1}{2} & 1 & \nicefrac{1}{2} & 1 \\
  \midrule
  $4$ & 216 & 6861 & 184 & 2621 & 0.99 & 0.942 \\
  $6$ & 5912 & 5800829 & 3732 & 473717 & 0.987 & 0.933 \\
  $8$ & 350872 & 47067992003 & 108534 & \textemdash & 0.982 & \textemdash \\
  $10$ & 37403668 & \textemdash & 6042418 & \textemdash & 0.982 & \textemdash \\
  $12$ & 10103561614 & \textemdash & \textemdash & \textemdash & \textemdash & \textemdash \\
  \bottomrule
\end{tabular}
  \caption{Number of fracton-free states, disconnected sectors and fragmentation coefficient from \cref{eq:fragmentation_coefficient} for spin-1/2 and spin-1 on $L\times L$ lattices. Dashes indicate data that was infeasible to obtain due to memory requirements. }
  \label{tab:sectors}
\end{table}

\section{Dynamically disconnected diagonal stripe sectors}\label{sec:relation_staircase}
Here we show that $90^\circ$ rotated versions of the diagonal stripe state lie in dynamically disconnected sectors under the action of $\mathcal{F}_{\sublO}$ and $\mathcal{F}_{\sublO}^\dagger$. This property explains the broken $90^\circ$ rotation symmetry of the spin structure factor in Fig.~5(a) of the main text and shows that this asymmetry is not a consequence of spontaneous lattice symmetry breaking, but rather due to the asymmetry of the diagonal stripe state itself.

In Fig.~\ref{fig:staircase_relation}(a) we show a $90^\circ$ rotated version of the diagonal stripe state in Fig.~3(a) of the main text. For the system to tunnel between both states, the $z$-components of the spins have to change by the amount illustrated in Fig.~\ref{fig:staircase_relation}(b), where $\Delta S^z_\text{diag}$ is the difference between the diagonal stripe state in Fig.~3(a) of the main text and the one in Fig.~\ref{fig:staircase_relation}(a). In a tunneling process this difference has to be generated by repeated actions of $\mathcal{F}_{\sublO}$ (or $\mathcal{F}_{\sublO}^\dagger$) on the different eight-site clusters $\sublO$. As an example, the change $\Delta S^z$ from $\mathcal{F}_{\sublO}$ for one particular eight-site cluster $\sublO$ is illustrated in Fig.~\ref{fig:staircase_relation}(c). The problem of describing this tunneling process with $\mathcal{F}_{\sublO}$ can be considered as a linear algebra problem that consists of finding a linear combination of the eight-site changes $\Delta S^z$ in Fig.~\ref{fig:staircase_relation}(c) that yields $\Delta S^z_\text{diag}$ in Fig.~\ref{fig:staircase_relation}(b). It needs to be taken into account that the $N_{\textrm{sites}}/2$ fluctuators $\mathcal{F}_{\sublO}$ on all clusters $\sublO$ are not linearly independent but have rank $N_{\textrm{sites}}/2-1$, as discussed in detail in Sec. II.B in Ref.~\cite{Niggemann2025a}. Therefore, to find a unique solution, one of the fluctuators $\mathcal{F}_{\sublO}$ needs to be omitted to ensure that the other fluctuators $\mathcal{F}_{\sublO}$ form a set of basis vectors for the local moves in the constrained subspace. Here, without loss of generality, we have omitted the fluctuator in row 1, column 2 (when counted from the bottom left corner) of the lattice array in Fig.~\ref{fig:staircase_relation}. The solution to express $\Delta S^z_\text{diag}$ via local fluctuators, obtained by simple matrix inversion, is illustrated in Fig.~\ref{fig:staircase_relation}(d) where the numbers $a$ on the sublattice 2 sites $\sublO$ correspond to the multiplicities $(\mathcal{F}_{\sublO})^a$ with which the integer spin changes $\Delta S^z$ in the eight-site clusters have to be applied. Importantly, these multiplicities contain {\it non-integer} $a=\pm1/2$ values, showing that fractional applications of $\mathcal{F}_{\sublO}$ would be needed to tunnel between both states. In other words, no sequential applications of $\mathcal{F}_{\sublO}$ exist that can realize this process and, consequently, both diagonal stripe states lie in different Hilbert space sectors. This result is independent of the choice of fluctuator $\mathcal{F}_{\sublO}$ that is omitted in the set of basis vectors.

We note that in addition to $90^\circ$ rotated versions of diagonal stripe states, also time-reversed ($S^z\rightarrow -S^z$) versions exist. However, time-reversed partners of diagonal stripe states are found to be in the {\it same} Hilbert space sector, which implies that the diagonal stripe order at $\mu<0.8J'$ in Fig.~4 of the main text is associated with spontaneous time-reversal symmetry breaking.

\section{Maximally flippable states in the \texorpdfstring{$6\times6$}{6×6} sector}
\label{app:6x6MF}
\begin{figure}
    \centering
    \includegraphics[width=\linewidth]{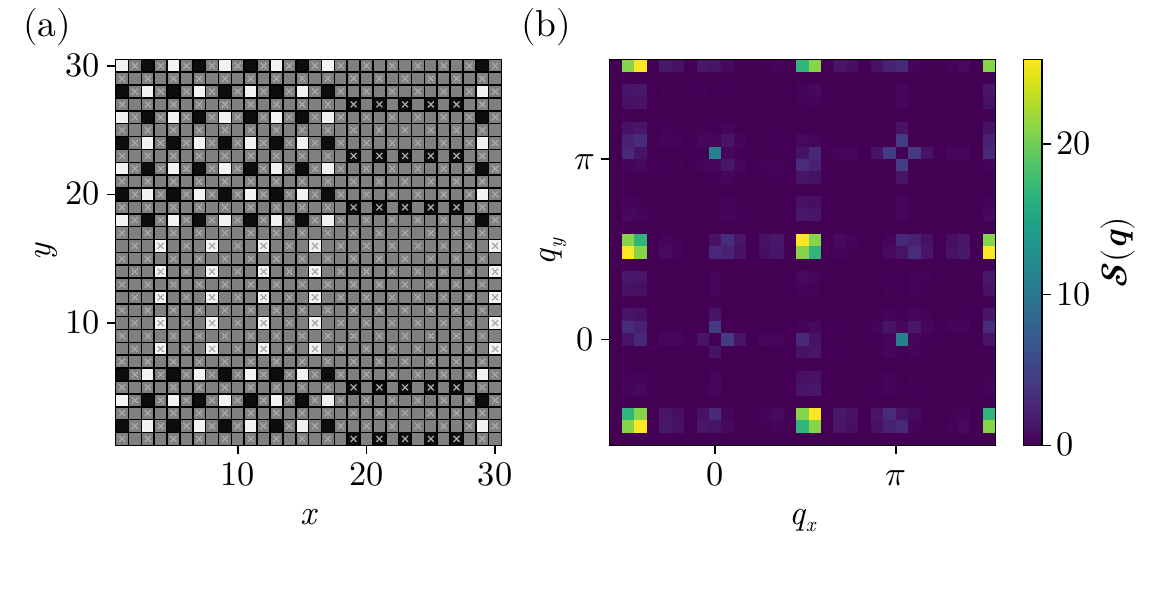}
    \caption{(a) One of the maximally flippable states found in the $6\times6$ sector and (b) its corresponding spin structure factor $\mathcal{S}({\bm q})$.}
    \label{fig:6x6_MF}
\end{figure}

The observation of broken ergodicity at low $\mu$ in the sector of the $6\times6$ state in Fig.~3(c) of the main text is related to the system's inability to exhibit $4\times 4$ order: While the $6\times6$ state has a large number of flippable clusters, numerical results indicate that other, more complicated, configurations exist in this sector with an even larger number. These configurations, such as the one shown in \cref{fig:6x6_MF}(a), and configurations related by lattice symmetries, are separated by large energetic barriers for small $\mu$, leading to exceedingly large autocorrelation times. At larger values of $\mu$, energetic barriers are lowered, while fluctuations between larger networks of spin configurations become favorable, allowing for ergodic quantum tunneling within the entire sector.
In order to find this configuration, we initialized 3840 independent GFMC simulations at $\mu=0$ with a single walker and recorded each configuration encountered. To confirm that the correct solution was found, we repeated this procedure five times, finding equivalent configurations (up to lattice translations). The result is shown in \cref{fig:6x6_MF}(a), featuring a phase separation effect with a large domain with a $4\times4$ order.  
By computing $\mathcal{S}(\bm{q})$ for this configuration [\cref{fig:6x6_MF}(b)], we may explain the features at $\bm{q} = (\pi/L,\pi/L)$ and symmetry-related points in Fig.~6(a) of the main text, as a result of the large size of the magnetic domains. On the other hand, this configuration still shows a strong peak at $\bm{q} = (\pi,\pi)$ which is however difficult to resolve in Fig.~6(a) of the main text due to the considerable uncertainties resulting from the broken ergodicity.

\section{Details on the effective rank-2 U(1) field theory}\label{sec:field_theory_details}
    Our Maxwellian field theory in Eq.~(9) of the main text is derived by expressing the spin flip operators in $\mathcal{H}_2$ in terms of rotor variables $S_i^\pm =\sqrt{2}e^{\pm \imath A^\alpha_i}$ where $A_i^\alpha$ takes the role of a generalized `vector' potential. The operators $A_i^\alpha$ (which follow the convention that $\alpha=xy$ when $i$ is on sublattice 1 and $\alpha=xx$ when $i$ is on sublattice 2) are the components of a trace-free and symmetric matrix-valued field, i.e., $A^{xx}_i=-A^{yy}_i$ and $A^{xy}_i=A^{yx}_i$. Furthermore, $A_i^\alpha$ is compact, i.e., its eigenvalues lie in the interval $[0,2\pi]$.
    The $z$-components of the spins $S_i^z$ take the role of a conjugate integer-valued matrix electric field $S_i^z=E_i^\alpha$ with $[A_i^\alpha,E_i^\alpha]=\imath$. Note that when $i$ is on sublattice 1 (2), we also use the notation $A_i^{xy}\equiv A^{xy}_{\sublX}$ ($A_i^{xx}\equiv A^{xx}_{\sublO}$) and equivalently for $E_i^\alpha$. With these definitions $\mathcal{H}_2$ has the form
    \begin{equation}
    \mathcal{H}_2\sim -\sum_{\sublO}\cos B_{\sublO}\sim+\sum_{\sublO}B_{\sublO}^2\label{eq:b_expansion}
    \end{equation}
    where ${B_{\sublO}}$ is defined for each cluster $\sublO$ as given in Eq.~(8) of the main text and the rightmost expression in Eq.~(18) of the main text represents the expansion of $\mathcal{H}_2$ in lowest non-trivial order in $B_{\sublO}$. Furthermore, in order to suppress electric field states $E_i^\alpha\notin \{-1,0,1\}$, a term $\sim \sum_{i} (E^\alpha_i)^2$ in the field theory is required. Together with the RK potential $\sim\sum_{\sublO}\mathcal{N}^2_{\sublO}$ this yields the effective field theory in Eq.~(9) of the main text. An important property of this theory that distinguishes it from conventional U(1) gauge theories is the absence of electromagnetic duality. This is already evident from the different properties of the electric and magnetic fields where the former is a matrix while the latter is a one-component object.

    Note that our mapping to an effective field theory could, in principle, also be applied to the spin-1/2 spiderweb model. This would require a change of the (irrelevant) prefactor $\sqrt{2}$ in Eq.~(6) of the main text and a redefinition of the electric field to ensure that it only assumes integer values. The resulting field theory, known e.g.\ from spin-1/2 quantum spin ice, is called `frustrated'~\cite{Hermele2004} due to the lack of a well-defined vacuum of electric fields. On the other hand, our field theory for a spin-1 system has the advantage of a unique vacuum state corresponding to the homogeneous $S_i^z=0$ configuration.
    
    The field theory has a local gauge freedom that follows from the rank-2 Gauss' law and amounts to the invariance under the operation~\cite{Savary-2017}
    \begin{equation}\label{eq:gauge_op}
    U(f_{\sublX})=\exp(\imath \sum_{\sublX} f_{\sublX}\mathcal{C}_{\sublX})
    \end{equation}
    where $f_{\sublX}$ is an arbitrary function defined for each cluster $\sublX$ or, stated differently, $f_{\sublX}$ is located on sublattice 1. To see this, we use
    \begin{equation}
    \mathcal{C}_{\sublX}=E^{xx}_{\sublX_1}+E^{xy}_{\sublX_2}-E^{xx}_{\sublX_3}-E^{xy}_{\sublX_4}+E^{xx}_{\sublX_5}+E^{xy}_{\sublX_6}-E^{xx}_{\sublX_7}-E^{xy}_{\sublX_8}
    \end{equation}
    and rearrange the sum in the exponent of Eq.~(\ref{eq:gauge_op}), yielding
    \begin{align}
    U(f_{\sublX})=&\exp[\imath \sum_{\sublX} E^{xy}_{\sublX}(f_{\sublX_2}-f_{\sublX_4}+f_{\sublX_6}-f_{\sublX_8})]\times\notag\\
    &\exp[\imath \sum_{\sublO} E^{xx}_{\sublO}(f_{\sublO_1}-f_{\sublO_3}+f_{\sublO_5}-f_{\sublO_7})].\label{eq:gauge_op2}
    \end{align}
    Here, the labels $\sublX_a$ and $\sublO_a$ use the convention explained below Eq.~(2) of the main text where, e.g., $f_{\sublO_1}$ is a site to the right of the center of a $\sublO$ cluster, such that $f_{\sublO_1}$ is still defined on sublattice 1 (center of a $\sublX$ cluster). Since $\exp (\imath \theta E_i^\alpha )$ with $\theta\in\mathds R$ shifts $A_i^\alpha\rightarrow A_i^\alpha+\theta$, the operation in Eq.~(\ref{eq:gauge_op2}) can be written as
    \begin{align}
    A^{xy}_{\sublX}&\rightarrow A^{xy}_{\sublX}+f_{\sublX_2}-f_{\sublX_4}+f_{\sublX_6}-f_{\sublX_8},\notag\\
    A^{xx}_{\sublO}&\rightarrow A^{xx}_{\sublO}+f_{\sublO_1}-f_{\sublO_3}+f_{\sublO_5}-f_{\sublO_7}.\label{eq:gauge_op3}
    \end{align}
    By simple bookkeeping of all terms it can be checked that Eq.~(8) of the main text is indeed invariant under this transformation.
    \begin{table}
    \begin{tabular}{|c|c|}
    \hline
    Quantities on sublattice 1 & Quantities on sublattice 2\\
    ($=$ center of $\sublX$ clusters): &($=$ center of $\sublO$ clusters):\\\hline
    $\mathcal{C}$ (constraint) & $\mathcal{F}$ (fluctuator)\\
    $\rho$ (fractons) & $B$ (magnetic field)\\
    $A^{xy}$, $E^{xy}$  & $A^{xx}$, $E^{xx}$\\
    (fields of U(1) theory) & (fields of U(1) theory)\\
    $f$ (gauge transformation)&$\mathcal{N}^2$ (RK potential)\\
    
    \hline
    \end{tabular}
    \caption{Definitions of the two sublattices of the spiderweb model and location of different quantities.}\label{tab1}
    \end{table}
    
    In Table~\ref{tab1} we list all the quantities occurring in the spiderweb model and in the effective field theory and we specify the sublattice on which they are defined.
    
    The gauge transformation in Eq.~(\ref{eq:gauge_op3}) becomes more transparent in a continuum description where it reads
    \begin{align}
    A^{xy}&\rightarrow A^{xy}+4\partial_x \partial_y f,\notag\\
    A^{xx}&\rightarrow A^{xx}+(\partial_x^2-\partial_y^2)f.\label{eq:gauge_op4}
    \end{align}
    Using the continuum definition of the magnetic field in Eq.~(8) of the main text,
    \begin{equation}\label{eq:bcontinuum}
    B=-4\partial_x\partial_y A^{xx}+(\partial_x^2-\partial_y^2)A^{xy},
    \end{equation}
    the invariance of $B$ under the gauge transformation in Eq.~(\ref{eq:gauge_op4}) is immediately obvious. The special property of Eq.~(\ref{eq:bcontinuum}) is that $B$ is constructed from {\it second} derivatives of $A^{xx}$ and $A^{xy}$. This is in contrast to a three-dimensional scalar charge rank-1 or rank-2 U(1) gauge theory where one derivative is sufficient to construct a gauge invariant magnetic field~\cite{Pretko2017_1}. The relation between $B$ and $A^{\mu\nu}$ in Eq.~(\ref{eq:bcontinuum}) also makes the lack of Lorentz-invariance of the effective field theory apparent. To see this, we consider the system's Lagrangian $\mathcal{L}$ which contains a term $\sim(\partial_t A^{\mu\nu})^2$ with a first derivative in time describing the electric field contribution $(E^{\mu\nu})^2$. Furthermore, $\mathcal{L}$ contains a term $\sim B^2$ which, according to Eq.~(\ref{eq:bcontinuum}), contains second spatial derivatives. This unequal treatment of space and time is incompatible with Lorentz invariance.
    
    Returning to the lattice version of our rank-2 U(1) gauge theory [see Eq.~(9) of the main text], this Hamiltonian describes a Gaussian theory whose eigenmodes can be calculated analytically. To this end we rewrite $A_i^\alpha$ and $E_i^\alpha$ in terms of bosonic operators $a_i^\alpha$, $(a_i^\alpha)^\dagger$,
    \begin{equation}\label{eq:bosons}
    A_i^\alpha=\frac{1}{\sqrt{2}}\left[a_i^\alpha+(a_i^\alpha)^\dagger\right]\;,\quad E_i^\alpha=\frac{1}{\sqrt{2}\imath}\left[a_i^\alpha-(a_i^\alpha)^\dagger\right]\;,
    \end{equation}
    which fulfill the standard commutation relation $[a_i^\alpha,(a_i^\alpha)^\dagger]=1$. Furthermore, we Fourier-transform the bosonic operators using
    \begin{equation}
    a_i^\alpha=\sqrt{\frac{2}{N_{\textrm{sites}}}}\sum_{\bm q} e^{\imath {\bm q}\cdot{\bm r}_i}a^\alpha({\bm q})\;.
    \end{equation}
    Here, ${\bm r}_i$ is the real-space position of site $i$ and the sum only includes momenta ${\bm q}$ in the first Brillouin zone. Note that our model consists of two inequivalent sublattices such that the first Brillouin zone can, for example, be chosen as $q_x\in[-\pi/2,\pi/2]$, $q_y\in[-\pi,\pi]$. With these definitions, $\mathcal{H}_\text{eff}$ from Eq.~(9) of the main text can be written as
    \begin{equation}
    \mathcal{H}_\text{eff}=\sum_{\bm q} A^\dagger({\bm q}) H_\text{eff}A({\bm q})\;,
    \end{equation}
    where $A({\bm q})$ is a four-component operator
    \begin{equation}
    A({\bm q})=(a^{xy}({\bm q}),a^{xx}({\bm q}),\left[a^{xy}(-{\bm q})\right]^\dagger,\left[a^{xx}(-{\bm q})\right]^\dagger)
    \end{equation}
    and the $4\times 4$ matrix $H_\text{eff}$ is given by
    \begin{equation}
    H_\text{eff}=K V_+^T V_+ +W V_-^T V_- +\frac{U}{4}\left(\begin{array}{cccc}1&0&-1&0 \\ 0&1&0&-1 \\ -1&0&1&0 \\ 0&-1&0&1\end{array}\right)
    \end{equation}
    with
    \begin{equation}
    V_\pm=(c_x-c_y,2s_x s_y,\pm(c_x-c_y),\pm2s_x s_y)
    \end{equation}
    and $c_\mu=\cos q_\mu$, $s_\mu=\sin q_\mu$ with $\mu=x,y$.
    
    According to the standard procedure of a Bogoliubov transformation, to solve the model, we need to rewrite $\mathcal{H}_\text{eff}$ in terms of Bose operators $C({\bm q})$ for the eigenmodes,
    \begin{equation}
    C({\bm q})=(c_1({\bm q}),c_2({\bm q}),c_1^\dagger({\bm q}),c_2^\dagger({\bm q}))
    \end{equation}
    where $C({\bm q})=\mathcal{J}(\bm q) A(\bm q)$ such that $(\mathcal{J}^\dagger)^{-1}H_\text{eff}\mathcal{J}^{-1}$ is diagonal. Note that, to preserve the bosonic commutation relations of $C({\bm q})$, the $4\times 4$ matrix $\mathcal{J}({\bm q})$ needs to be a {\it paraunitary} transformation that satisfies $\mathcal{J}^\dagger g \mathcal{J}=g$ with $g=\text{diag}(1,1,-1,-1)$~\cite{Colpa1978}. A subtlety arises because of the gauge invariance of $\mathcal{H}_\text{eff}$ which manifests in a bosonic zero mode. Since a paraunitary transformation $\mathcal{J}({\bm q})$ is not defined in this case, we introduce an additional term in the Hamiltonian of the form $2d\sum_i(A_i^\alpha)^2$ which breaks gauge invariance and which allows us to explicitly calculate $\mathcal{J}({\bm q})$. The gauge-invariant limit $d\rightarrow 0$ can be recovered at the end of the calculation. The transformation matrix $\mathcal{J}({\bm q})$ is found to be
    \begin{align}
    \mathcal{J}({\bm q})&=\frac{1}{\sqrt{8(L_1^2+L_2^2)}}\times\notag\\
    &\times\left(\begin{array}{cccc}
    L_1\xi_+&
    L_2\xi_+&
    -L_1\xi_-&
    -L_2\xi_-\\
    -L_2\lambda_+&
    L_1\lambda_+&
    L_2\lambda_-&
    -L_1\lambda_-\\
    -L_1\xi_-&
    -L_2\xi_-&
    L_1\xi_+&
    L_2\xi_+\\
    L_2\lambda_-&
    -L_1\lambda_-&
    -L_2\lambda_+&
    L_1\lambda_+
    \end{array}\right) 
    \end{align}
    with
    \begin{align}
    \xi_\pm&=\left(\frac{U}{d}\right)^{1/4}\pm 2\left(\frac{d}{U}\right)^{1/4},\\
    \lambda_\pm&=\sqrt{2}\left[\left(\frac{\eta_2}{\eta_1}\right)^{1/4}\pm \left(\frac{\eta_1}{\eta_2}\right)^{1/4}\right],\\
    \eta_1&=d+K\left[(c_x-c_y)^2+4s_x^2 s_y^2\right],\\
    \eta_2&=\frac{U}{4}+W\left[(c_x-c_y)^2+4s_x^2 s_y^2\right],
    \end{align}
    and $L_1$, $L_2$ are the components of the \emph{constraint vector}~\cite{Niggemann2025a}, given by
    \begin{equation}
    L_1(\bm q)=-4 s_x s_y\;,\;
    L_2(\bm q)=2(c_x-c_y)\;.\label{eq:constraint_L2}
\end{equation}
    Diagonalizing $H_\text{eff}$ with $\mathcal{J}$ leads in the limit $d\rightarrow 0$ to a zero mode due to the system's gauge freedom and a single photon mode with the dispersion
    \begin{align}\label{eq:full_photon_disp}
    \omega(\bm q)&=2\sqrt{\eta_1\eta_2}\notag\\
    &=2\sqrt{K\left[(c_x-c_y)^2+4s_x^2 s_y^2\right]}\times\notag\\
    &\times\sqrt{\frac{U}{4}+W\left[(c_x-c_y)^2+4s_x^2 s_y^2\right]}.
    \end{align}
    This photon dispersion is gapless at $\bm q=(0,0)$ and $\bm q=(\pi,\pi)$. An expansion of $\omega(\bm q)$ around these two points yields for $U\neq0$ in lowest non-vanishing order
    \begin{equation}
    \omega(\bm q)\approx\sqrt{\frac{KU}{4}}\sqrt{q_x^4+14 q_x^2 q_y^2+q_y^4}\;.
    \end{equation}
    This function is quadratic in any radial direction away from the gapless points, however, it does not have a continuous rotation symmetry around these points, as discussed in Sec.~II.A in Ref.~\cite{Niggemann2025a}. On the other hand, exactly at the RK-point $U=0$, the photon dispersion becomes quartic at long wavelengths,
    \begin{equation}
    \omega(\bm q)\approx\sqrt{\frac{KW}{4}}\left(q_x^4+14 q_x^2 q_y^2+q_y^4\right)\;.
    \end{equation}
    Another prediction of the field theory is the spin structure factor
    \begin{equation}
    \mathcal{S}(\bm q)=\frac{1}{N_{\textrm{sites}}}\sum_{i,j}\langle S_i^z S_j^z\rangle e^{\imath {\bm q}\cdot({\bm r}_i-{\bm r}_j)},
    \end{equation}
    which can be obtained by expressing the spin operators $S_i^z$ in terms of $a_i^\alpha$, $(a_i^\alpha)^\dagger$ bosons [Eq.~(7) of the main text and Eq.~(\ref{eq:bosons})], transforming them into the eigenbasis of $C$ bosons using the matrix $\mathcal{J}$, and exploiting that the ground state is free of any photon excitations. For $d\rightarrow0$ this yields 
    \begin{align}\label{eq:ssf_analytical}
    &\mathcal{S}(\bm q)=\sqrt{\frac{\eta_1}{\eta_2}}\frac{(L_1-L_2)^2}{L_1^2+L_2^2}\notag\\
    &=\frac{\sqrt{K}(c_x-c_y+2s_xs_y)^2}{\sqrt{(c_x-c_y)^2+4s_x^2 s_y^2}\sqrt{\frac{U}{4}+W\left[(c_x-c_y)^2+4s_x^2 s_y^2\right]}}.
    \end{align}
    Note that for fitting this function to numerical results, we need to define only two truly independent fitting parameters $(A,r)$ via $(K,W,U) = (4A^2,1,r)$, where $A$ only modifies the structure factor by a global scaling, which cannot be fixed by sum rules for spin 1, and $r$ tunes the relative strength of the RK potential. 
    In the RK limit $U\rightarrow0$ where $\eta_1/\eta_2=K/W$ is a constant, this expression becomes (up to a prefactor) identical to the classical spin structure factor in Gaussian approximation given by $\mathcal{S}_\textrm{class}({\bm q}) = \frac{(c_x-c_y+2s_x s_y)^2}{(c_x-c_y)^2+4 s_x^2 s_y^2}$~\cite{Niggemann2025a}. This is expected because at the RK point a ground state can be constructed by an equal weight superposition of all $S_i^z$ basis states in a given Hilbert space sector, similar to a classical (non-coherent) superposition. Furthermore, at $W=0$ when $\eta_2=U/4$ is a constant, the spin structure factor in Eq.~(\ref{eq:ssf_analytical}) corresponds to the classical result, multiplied by the photon dispersion $\omega(\bm q)$ which suppresses the fourfold pinch points around their center. The interpolation between both limits is determined by the term $\sqrt{U/4+W\left[(c_x-c_y)^2+4s_x^2 s_y^2\right]}$ in the denominator of Eq.~(\ref{eq:ssf_analytical}). For finite $W>0$ there is a threshold momentum $q_c$ (which decreases with increasing $W$) above which the $W$-term dominates and the spin structure factor resembles the classical one. On the other hand, for $q\lesssim q_c$ the $U$-term dominates and the spin structure factor is suppressed.

    The photon dispersion $\omega(\bm q)$ and the spin structure factor $\mathcal{S}(\bm q)$ of the $90^\circ$ rotation symmetry broken field theory with the additional term in Eq.~(12) of the main text follow straightforwardly from Eq.~(\ref{eq:full_photon_disp}) and Eq.~(\ref{eq:ssf_analytical}) by the replacement
    \begin{equation}
    U\rightarrow U[1+2p(1+\cos(q_x-q_y))],
    \end{equation}
    where $p=U'/U$.
    As pointed out in the main text, we define the independent parameters $(A,r,p)$ by the relation $(K,W,U,U') = (4A^2,1,r,pr)$, for fitting the spin structure factor of the asymmetric field theory.

\section{Real-space correlations of the rank-2 U(1) field theory}\label{sec:real_space_correlations}
Here, we calculate the real-space spin correlations $C(\bm r)$ of the effective rank-2 U(1) field theory by analytically Fourier-transforming the spin structure factor derived in Eq.~(\ref{eq:ssf_analytical}) of the last section,
\begin{equation}
    C({\bm r})=\int\frac{d^2{\bm q}}{(2\pi)^2}\mathcal{S}({\bm q}) e^{\imath{\bm q}\cdot{\bm r}}.\label{eq:def_correlation}
\end{equation}
We are specifically interested in the type of spatial decay of the correlations at long distances, i.e., in the continuum limit at ${\bm q}\rightarrow 0$. For any finite $W$ there is a range of small momenta where the first term $U/4$ in the second square root of the denominator in Eq.~(\ref{eq:ssf_analytical}) dominates over the second term $\sim W$. In this regime $\mathcal{S}({\bm q})$ becomes
\begin{equation}
\mathcal{S}({\bm q})=\sqrt{\frac{K}{U}}\frac{(L_1-L_2)^2}{\sqrt{L_1^2+L_2^2}}.\label{eq:ssf_small_q}
\end{equation}
To make further progress by analytical calculations, we slightly modify the field theory of the last section and replace the first component of the constraint vector in Eq.~(\ref{eq:constraint_L2}) by
\begin{equation}
L_1({\bm q})=-4s_x s_y\longrightarrow -2 s_x s_y.\label{eq:L_replacement}
\end{equation}
This replacement is motivated by the fact that, in the limit ${\bm q}\rightarrow 0$, the norm of the constraint vector now depends solely on $q$ and is independent of the polar angle,
\begin{equation}
L_1^2+L_2^2= q^4\;\text{at small $q$},
\end{equation}
making the analytical derivation of the polar angle integration in Eq.~(\ref{eq:def_correlation}) possible. On the other hand, the radial dependence of $\mathcal{S}({\bm q})$, and thus the decay behavior of real-space correlations, which is our primary interest here, remains unaffected. From a physical perspective, this replacement makes the photon dispersion rotation invariant at small momenta.

Inserting Eq.~(\ref{eq:ssf_small_q}) into Eq.~(\ref{eq:def_correlation}) and expanding the constraint vector in lowest order in $q$ one obtains
\begin{align}
C({\bm r})&\sim\int\frac{d^2{\bm q}}{(2\pi)^2}(q_x^2-q_y^2+2q_x q_y)^2\frac{e^{\imath{\bm q}\cdot{\bm r}}}{q^2}\notag\\
&=(\partial_x^2-\partial_y^2+2\partial_x\partial_y)^2\int\frac{d^2{\bm q}}{(2\pi)^2}\frac{e^{\imath{\bm q}\cdot{\bm r}}}{q^2}\notag\\
&\equiv (\partial_x^2-\partial_y^2+2\partial_x\partial_y)^2 g({\bm r}).
\end{align}
The function $g({\bm r})$ defined in the last line obeys the relation
\begin{equation}
(\partial_x^2+\partial_y^2)g({\bm r})=-\int \frac{d^2{\bm q}}{(2\pi)^2}e^{\imath{\bm q}\cdot{\bm r}}=-\delta(\bm q).
\end{equation}
This equation can be easily integrated, yielding $g(\bm r)\sim \log(|\bm r|)$, which results in the real-space correlations
\begin{equation}
C(\bm r)\sim\frac{xy(-x^2+y^2)}{|\bm r|^8}.
\end{equation}
Most importantly, $C(\bm r)$ decays as $\sim \frac{1}{|\bm r|^4}$ along any radial direction.
 
\section{Further indications for a gapless higher-rank spin liquid}\label{sec:further_criteria}
The main text demonstrates that the phase diagram of the spin-1 spiderweb model displays a transition from a disordered, spin-liquid phase adiabatically connected to the solvable RK point to an ordered phase at finite $\mu_c$, which can differ between sectors. This spin liquid phase was found to be well described by a rank-2 U(1) quantum gauge theory.

\begin{figure}[h]
    \centering
    \includegraphics[width= \linewidth]{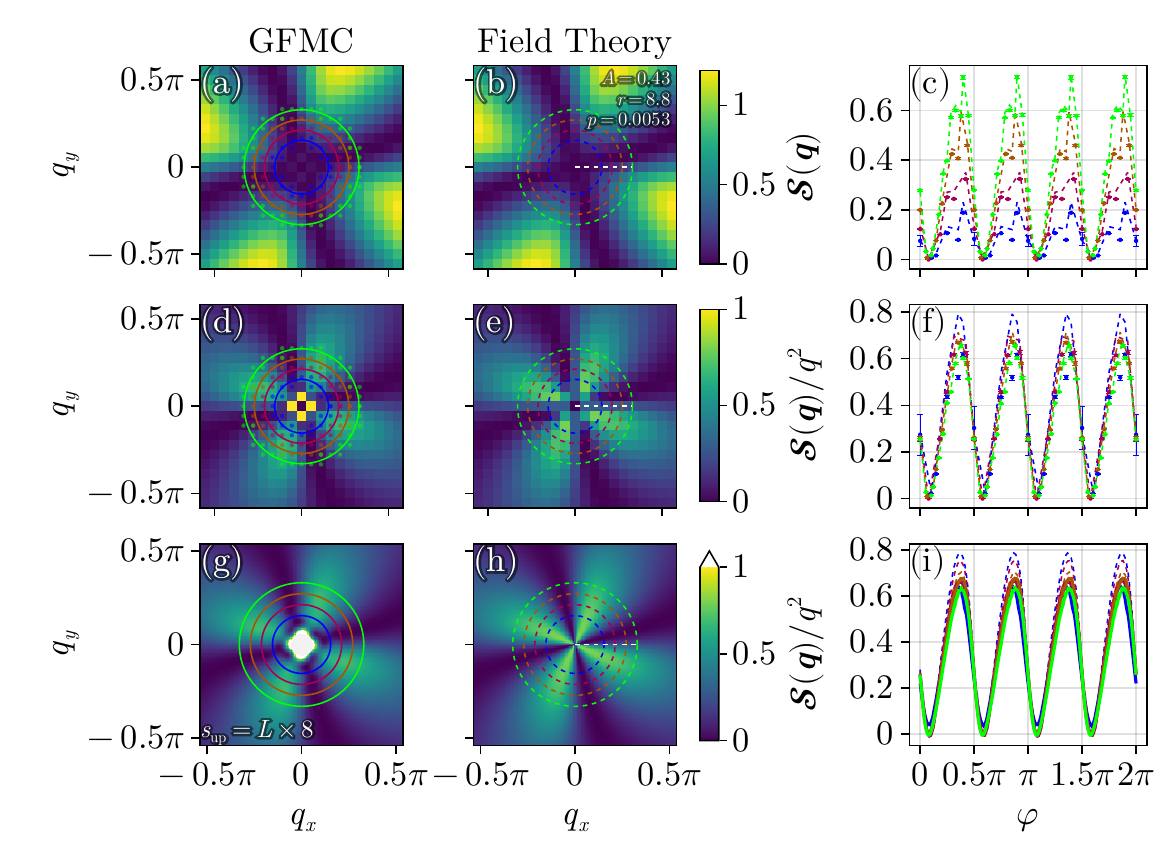}
    \caption{
      $\mathcal{S}(q,\varphi)$ close to the pinch point for $L=36$ at $\mu=0.8J'$ for GFMC (a,d,g) and field theory (b,e,h) in the $6\times6$ sector. Also shown are momentum discretized, circular paths with a starting point $\varphi=0$ indicated by a white dashed line. (c,f,i): Circular cuts of $\mathcal{S}(q,\varphi)$ for the paths shown.
      (d,e,f) shows the rescaled quantity $\mathcal{S}(q,\varphi)/q^2$. (g): Same data in (d), but upsampled by a factor of 8 using sinc interpolation. The radius of white area at ${\bm q} = (0,0)$ indicates the resolution limit $\sim 2\pi/L$ permitted by the finite system size, above which correlations are neglected.
      The field theory estimate (h) is not upsampled and has perfect resolution.
    }
    \label{fig:6x6Spin1_Sq_RadialCuts_6x6_mu08}
\end{figure}

Here, we further investigate the nature of the pinch point suppression, discussed in the main text and derived in the previous section.
Expressed in radial coordinates $q = | {\bm q}|$ and $\varphi$, close to the pinch point origin, i.e. at ${\bm q} = (0,0)$, the spin structure factor from the field theory in \cref{eq:ssf_analytical} is of the form
\begin{equation}
    \mathcal{S}(q,\varphi) \approx g(\varphi) q^2 \label{eq:Sq_expansion_pinchpoint}
\end{equation}
where $g(\varphi)$ is a function of only the angle $\varphi$. It is thus insightful to rescale the spin structure factor by $1/q^2$, as this should reveal the singular nature of the fourfold pinch point. The result is shown for the $6\times 6$ sector at $\mu=0.8J'$ in Fig.~\ref{fig:6x6Spin1_Sq_RadialCuts_6x6_mu08}.
For reference, panels (a-c) show a comparison of GFMC and field theory results for $\mathcal{S}(q,\varphi)$ without such rescaling for several circular paths around the pinch point.
After rescaling $\mathcal{S}(\mathbf{q})/q^2$, in both GFMC, Fig.~\ref{fig:6x6Spin1_Sq_RadialCuts_6x6_mu08}(d) and field theory, Fig.~\ref{fig:6x6Spin1_Sq_RadialCuts_6x6_mu08}(e), the fourfold pinch point becomes visible. Moreover, the curves showing the cuts along the circular paths in Fig.~\ref{fig:6x6Spin1_Sq_RadialCuts_6x6_mu08}(f) collapse onto a single curve, confirming the expected form in \cref{eq:Sq_expansion_pinchpoint}. 
In order to better visualize this collapse without artifacts from the finite momentum space resolution, we also show a sinc-interpolated version of the GFMC data for $\mathcal{S}(q,\varphi)$ in panels (g,i).
We find that this interpolation is the most natural, as it corresponds to a Fourier transform of an effectively enlarged system size, where long-distance correlations $\gtrsim L$ are neglected.
Aside from artifacts at the smallest momenta (which correspond to inaccuracies of the vanishingly small long-range correlations), we find that this reproduces the continuum field theory shown in panel (h) exceptionally well.

\begin{figure}[h]
    \centering
    \includegraphics[width= \linewidth]{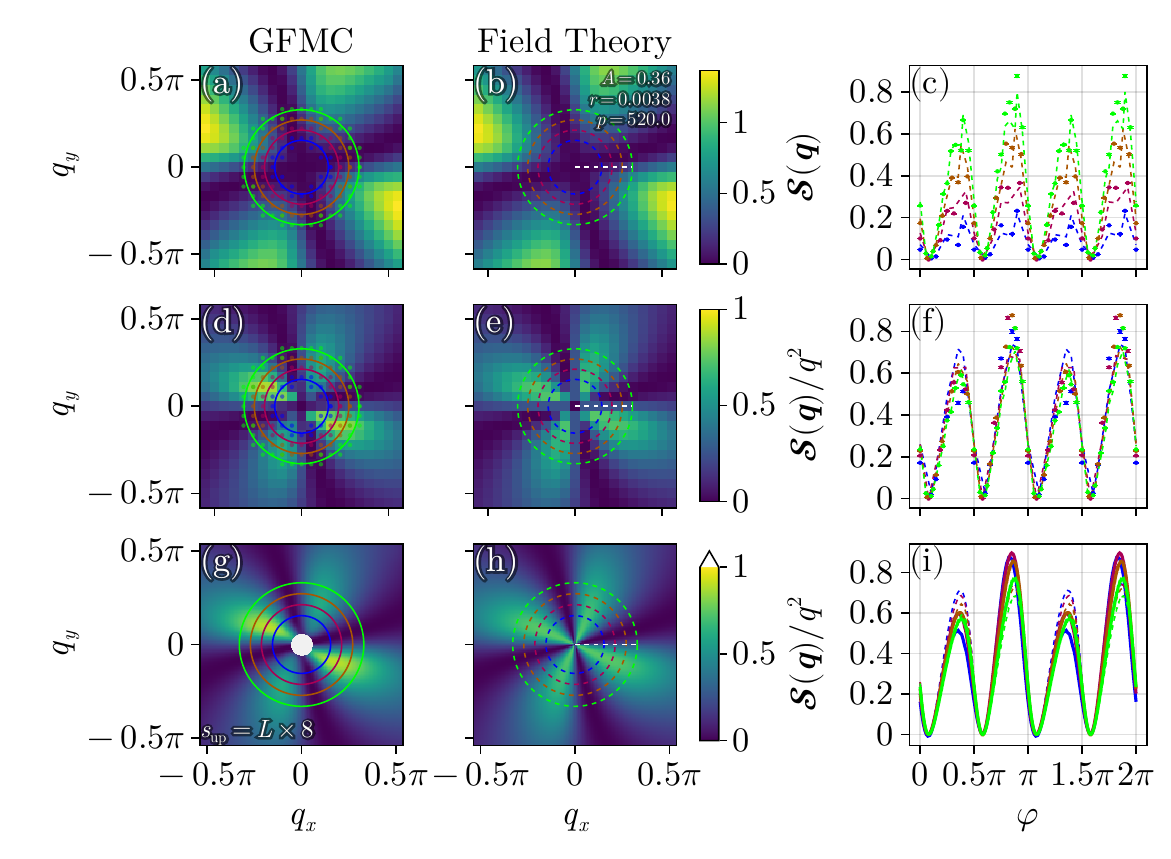}
    \caption{
    Same as in Fig.~\ref{fig:6x6Spin1_Sq_RadialCuts_6x6_mu08} but in the diagonal stripe sector for $\mu = 0.9J'$.
    }
    \label{fig:StairCaseSpin1_Sq_RadialCuts}
\end{figure}

\begin{figure}
    \centering
    \includegraphics[width=\linewidth]{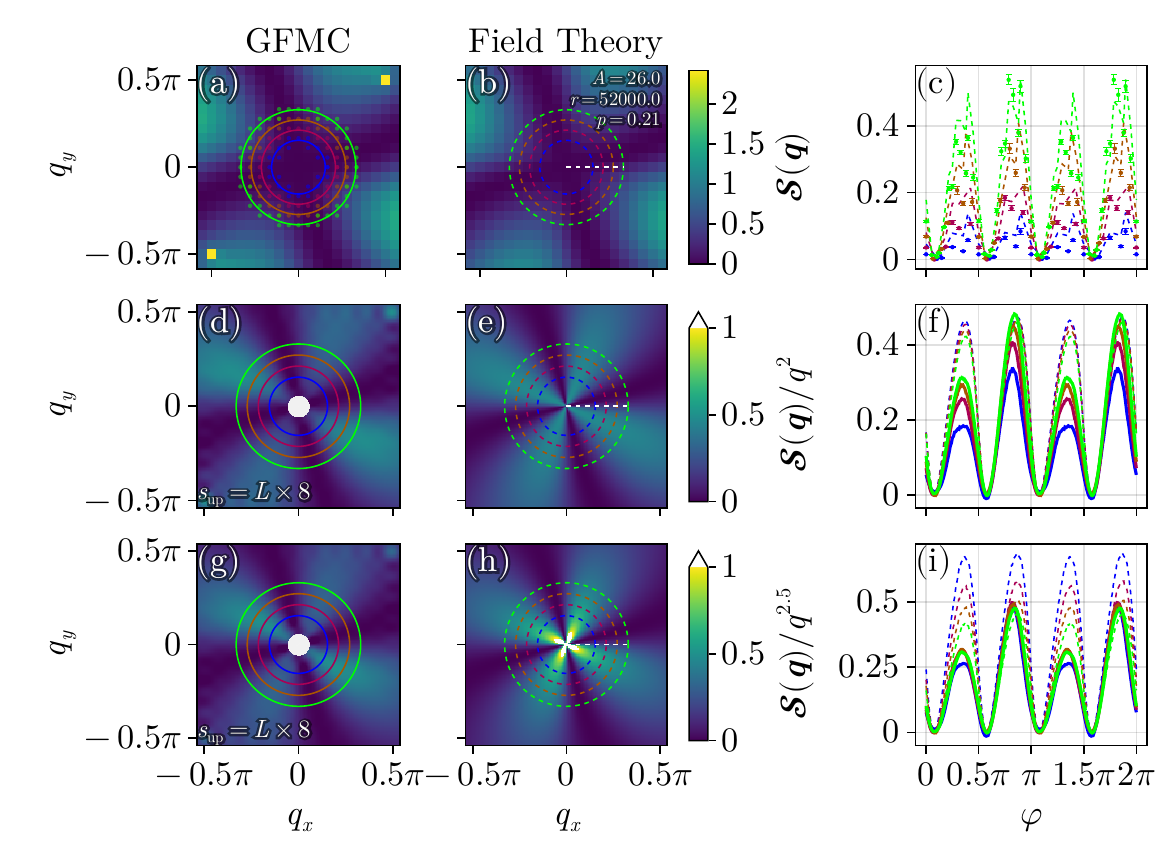}
    \caption{As in Fig.~\ref{fig:StairCaseSpin1_Sq_RadialCuts}, but in the ordered phase for $\mu=0.6J'$. Panels (g-i) show the collapse for a structure factor rescaled by a higher power $1/q^{2.5}$.}
    \label{fig:StaircaseSq_mu06}
\end{figure}

We also repeat this analysis for the diagonal stripe sector at $\mu=0.9J'$ and $\mu = 0.6J'$, respectively, in Figs.~\ref{fig:StairCaseSpin1_Sq_RadialCuts} and \ref{fig:StaircaseSq_mu06}. In the spin-liquid phase at $\mu=0.9J'$, the findings are largely equivalent, aside from a slightly enhanced asymmetry of the spin structure factor in the GFMC data.

The observation of a spin structure factor of the form in Eq.~(\ref{eq:Sq_expansion_pinchpoint}) changes in the ordered phase at $\mu=0.6J'$ of the diagonal stripe sector, where we find a much stronger suppression $\sim q^{5/2}$ in contrast to the best field-theory fit; see Fig.~\ref{fig:StaircaseSq_mu06}. We note that the precise exponent of the scaling is difficult to obtain to high accuracy, rendering an estimation of the phase boundaries using this feature rather difficult.

\begin{figure*}
    \centering
    \includegraphics[width=\linewidth]{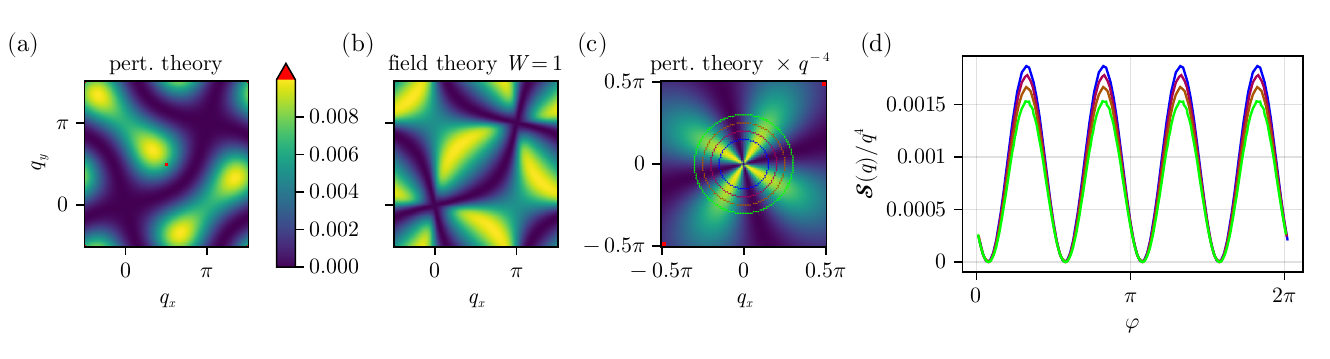}
    \caption{(a) Spin structure factor in the sector of the diagonal stripe state at $\mu\rightarrow-\infty$ computed in first order perturbation theory in $\mathcal{H}_2$. The color scale was truncated to not include the ordering peak so that the perturbative fluctuations are visible. (b) Prediction of the spin structure factor of the fractonic U(1) field theory of Eq.~(\ref{eq:ssf_analytical}) at $W=1$. (c) Same as (a) but rescaled by $\frac{1}{q^4}$ and plotted in a smaller part of the momentum space. (d) Spin structure factor in (c) plotted along the rings shown in (c).}
    \label{fig:first_order}
\end{figure*}

A qualitative explanation of this stronger suppression is obtained by considering the perturbative limit of the stripe ordered state, denoted $|\psi_\text{st}\rangle$, at $\mu \rightarrow -\infty$. Since $|\psi_\text{st}\rangle$ is an eigenstate of $\mathcal{H}_1$ and $\mathcal{H}_3$ and since $\mathcal{H}_2$ is much smaller than $\mathcal{H}_1+\mathcal{H}_3$ we treat the effect of $\mathcal{H}_2$ on $|\psi_\text{st}\rangle$ perturbatively. In first order in $\mathcal{H}_2$, the perturbed state takes the form
\begin{equation}\label{eq:first_order}
|\psi^{(1)}\rangle=|\psi_\text{st}\rangle-\frac{J'}{2\mu}\sum_{\sublO}\left(\mathcal{F}_{\sublO}+\mathcal{F}^\dagger_{\sublO}\right)|\psi_\text{st}\rangle.
\end{equation}
The spin structure factor $\mathcal{S}(\bm q)$ of $|\psi^{(1)}\rangle$ shown in Fig.~\ref{fig:first_order}(a) features magnetic Bragg peaks at $\bm q =\pm(\pi/2,\pi/2)$ indicative of diagonal stripe order. Furthermore, the spin structure factor exhibits a diffuse signal from the application of fluctuator moves $\mathcal{F}_{\sublO}$ and $\mathcal{F}_{\sublO}^\dagger$ on $|\psi_\text{st}\rangle$. This diffuse signal has the analytical form
\begin{equation}
\mathcal{S}_\text{diff}(\bm q)\sim [\cos(q_x) - \cos(q_y) + 2 \sin(q_x) \sin(q_y)]^2.
\end{equation}
Interestingly, $\mathcal{S}_\text{diff}(\bm q)$ also has a characteristic structure of nodal features and suppressed pinch points. However, its precise shape is different from the spin structure factor of our fractonic U(1) gauge theory shown for $W=1$ in Fig.~\ref{fig:first_order}(b) for comparison (see Eq.~(\ref{eq:ssf_analytical}) for the analytical expression). Specifically, as shown in Fig.~\ref{fig:first_order}(c) and (d), this first-order spin structure factor scales as $q^4$ for small momenta, explaining the observed suppression stronger than $q^2$ once magnetic long-range order sets in.

\end{appendices}

\end{document}